\documentclass[a4paper,11pt]{article}
\pdfoutput=1 

\usepackage{jheppub} 

\usepackage[T1]{fontenc} 
\usepackage{lineno}
\usepackage[figuresright]{rotating}
\usepackage{tabularx}
\usepackage{float}

\input{definitions}

\title{Study of $\Upsilon(10753)$ decays to $\pi^{+}\pi^{-}\Upsilon(nS)$ final states at \belletwo}

\collaboration{The Belle II Collaboration}
  \author{I.~Adachi\,\orcidlink{0000-0003-2287-0173},} 
  \author{L.~Aggarwal\,\orcidlink{0000-0002-0909-7537},} 
  \author{H.~Ahmed\,\orcidlink{0000-0003-3976-7498},} 
  \author{H.~Aihara\,\orcidlink{0000-0002-1907-5964},} 
  \author{N.~Akopov\,\orcidlink{0000-0002-4425-2096},} 
  \author{A.~Aloisio\,\orcidlink{0000-0002-3883-6693},} 
  \author{N.~Anh~Ky\,\orcidlink{0000-0003-0471-197X},} 
  \author{D.~M.~Asner\,\orcidlink{0000-0002-1586-5790},} 
  \author{H.~Atmacan\,\orcidlink{0000-0003-2435-501X},} 
  \author{T.~Aushev\,\orcidlink{0000-0002-6347-7055},} 
% ...

\abstract{We present an analysis of the process $e^{+}e^{-}\to\pi^{+}\pi^{-}\Upsilon(nS)$ (where $n$ = 1, 2, or 3) reconstructed in $19.6\rm$ \invfb of Belle II data during a special run of the SuperKEKB collider at four energy points near the peak of the $\Upsilon(10753)$ resonance.
By analyzing the mass distribution of the $\pi^+\pi^-\Upsilon(nS)$ system and the Born cross sections of the $e^{+}e^{-}\to\pi^{+}\pi^{-}\Upsilon(nS)$ process, we report the first observation of $\Upsilon(10753)$ decays to the $\pi^{+}\pi^{-}\Upsilon(1S)$ and $\pi^{+}\pi^{-}\Upsilon(2S)$ final states, and find no evidence for decays to $\pi^{+}\pi^{-}\Upsilon(3S)$.
Possible intermediate states in the $\pi^+\pi^-\Upsilon(1S,2S)$ transitions are also investigated, and no evidence for decays proceeding via the $\pi^\mp Z_b^\pm$ or $f_0(980)\Upsilon(nS)$ intermediate states is found.
We measure Born cross sections for the $e^{+}e^{-}\to\pi^{+}\pi^{-}\Upsilon(nS)$ process that, combined with results from Belle, 
obtain the mass and width of $\Upsilon(10753)$ to be $(10756.6\pm2.7\pm0.9)$ MeV/$c^2$ and $(29.0\pm8.8\pm1.2)$ MeV, respectively.
The relative ratios of the Born cross sections at the $\Upsilon(10753)$ resonance peak are also reported for the first time.
}

\begin{document}

\hyphenpenalty=10000
\tolerance=1000

\maketitle

\section{Introduction}
The Belle experiment observed a narrow enhancement in the cross section for the process $\epem\to\pi^{+}\pi^{-}\Upsilon(nS)$ $(n=1,2,3)$~\cite{exp_belle}.
The structure, with a mass and width of $M=(10753\pm6)~{\rm MeV}/c^2$ and $\Gamma=(36^{+18}_{-12})~\rm MeV$, respectively, is named $\Upsilon(10753)$~\cite{pdg}.
Several competing interpretations have been proposed for this structure, including a conventional bottomonium~\cite{Chen:2019uzm,Liang:2019geg,Li:2019qsg,Dong:2020tdw,Giron:2020qpb,vanBeveren:2020eis,Li:2021jjt,Bai:2022cfz,Husken:2022yik,Kher:2022gbz,Li:2022leg}, a hybrid~\cite{Brambilla:2019esw,TarrusCastella:2021pld}, or a tetraquark state~\cite{Ali:2019okl,Wang:2019veq,Bicudo:2020qhp,Bicudo:2022ihz}, but there is no definitive explanation so far. 
The Belle II experiment observed the process $\epem\to\omega\chi_{b1,2}(1P)$ at center-of-mass (c.m.)\ energies ($\sqrt{s}$) near $10.746$ GeV~\cite{PhysRevLett.130.091902}, confirming the existence of the $\Upsilon(10753)$ and identifying new decay channels of this state.
The ratio of the cross section of $\epem\to\omega\chi_{b1}(1P)$ to $\omega\chi_{b2}(1P)$ is $1.3\pm0.6$ at $\sqrt{s}=10.746$ GeV, which disagrees with the expectation for a pure $D$-wave bottomonium state, approximately 15~\cite{Guo:2014qra} and deviates from predictions for a $S$--$D$-mixed state, 0.18 - 0.22~\cite{Li:2021jjt}.
In particular, previous Belle measurements~\cite{exp_belle}, which uses Belle data only, are consistent with predictions in the $4S$-$3D$ mixing model~\cite{Bai:2022cfz}.
Further measurements of the properties and decay modes of the $\Upsilon(10753)$ are important to advance our understanding of its nature and test theoretical predictions.

In this paper we present an analysis of $\Upsilon(10753)\to\pipi\Upsilon(nS)$ using new, large samples of electron-positron collision data collected explicitly for this purpose by the \belletwo experiment. 
We reconstruct decays to $\pi^{+}\pi^{-}\Upsilon(nS)$ final states, with the $\Upsilon(nS)$ decaying to a $\mu^{+}\mu^{-}$ pair, at $\sqrt{s}$ in the $10.6$--$10.8$ \gev range. 
We measure and fit the Born cross sections ($\sigma_{\rm B}$) for these processes as a function of $\sqrt{s}$ to obtain the $\Upsilon(10753)$ mass and width. 
We search for intermediate states to study the internal decay dynamics (\emph{e.g.}, $\epem\to f_{0}(980)[\to\pi^{+}\pi^{-}]\Upsilon(nS)$) and exotic states ($\epem\to\pi^{\mp}Z_{b}(10610, 10650)^{\pm}[\to\pi^{\pm}\Upsilon(nS)$]), which may provide deeper insight into the possibility of an unconventional nature for the $\Upsilon(10753)$.

\section{Belle II detector and simulation}
The \belletwo detector~\cite{b2tdr, b2tip} operates at the SuperKEKB asymmetric-energy electron-positron collider~\cite{superkekb} at KEK. 
The \belletwo detector is a nearly $4\pi$ spectrometer consisting of silicon-based vertexing and drift-chamber tracking systems, Cherenkov-light particle identification detectors, and an electromagnetic calorimeter, situated within a superconducting solenoid providing a 1.5~T axial magnetic field.
The flux return of the solenoid is instrumented for $K^0_L$ and muon detection. 
The $z$ axis in the laboratory frame is collinear with the symmetry axis of the solenoid and nearly aligned with the electron-beam direction.

In addition to regular data taking at the peak of the $e^{+}e^{-}\to\Upsilon(4S)$ production cross section at $\sqrt{s}=10.58$ \gev,
in November 2021 SuperKEKB operated above the $\Upsilon(4S)$ resonance at $\sqrt{s}=10.653$, $10.701$, $10.746$, and $10.805$ \gev for studies of the $\Upsilon(10753)$. 
These energy points were selected to fill the gaps between previous collision energies studied by the Belle experiment in order to improve coverage of this region of interest. This analysis uses $3.5$, $1.6$, $9.8$, and $4.7$ \invfb of integrated luminosity at these points, respectively.

Simulated events are used for optimization of event selections, determination of reconstruction efficiencies, extraction of signal-resolution functions, and devising the fit models to extract the signals.
We generate $\Upsilon(10753)$ events with {\sc EvtGen}~\cite{evtgen}.
Initial-state radiation (ISR) at next-to-leading order accuracy in quantum electrodynamics is simulated with {\sc phokhara}~\cite{phokhara}.
Final-state radiation from stable charged particles is simulated using \texttt{PHOTOS}~\cite{PHOTOS}.
The process $e^{+}e^{-}\to\pi^{+}\pi^{-}\Upsilon(nS)$ is initially simulated assuming a phase-space model.
Detector simulation is performed with Geant4~\cite{AGOSTINELLI2003250}.
Reconstruction of events from simulated and collision data uses the Belle~II analysis software~\cite{basf2, basf2-zenodo}.
Additional simulated samples of low-multiplicity quantum electrodynamic processes, \emph{e.g.}, Bhabha scattering~\cite{babayga1,babayaga2,babayaga3,babayaga4,babayaga5}$\mu^{+}\mu^{-}(\gamma)$~\cite{babayga1,babayaga2,babayaga3,babayaga4,babayaga5}, ISR-produced hadron pairs~\cite{phokhara}, and four-track events with at least one lepton pair~\cite{aafh1,aafh2}, are used to check for contamination from possible backgrounds.

\section{Event selection}
Events are selected online by a hardware trigger that uses drift-chamber and calorimeter information with an efficiency greater than $97\%$ for events containing at least three tracks according to simulation~\cite{trigger}.
In the offline analysis, tracks reconstructed in the final state are required to originate from the vicinity of the interaction point (within $\SI{4}{\centi\meter}$ along the $z$ axis, and $\SI{2}{\centi\meter}$ in the radial direction) to remove beam-related backgrounds and incorrectly reconstructed tracks.
Tracks are required to be within the angular acceptance of the drift chamber, \emph{i.e.}, the polar angle with respect to the $z$ axis, $\theta$, should satisfy $-0.866<\cos\theta<0.9563$. 
We require that events contain four or five tracks to reduce backgrounds while allowing for increased efficiency for signal events with an additional track, which may be wrongly reconstructed from the detector noise.

We reconstruct the $\Upsilon(nS)$ candidate decaying to a pair of oppositely charged particles each with a momentum in the $\epem$ c.m.\ frame in the range $4.2<p^{*}(\mu)<5.35$ GeV/$c$, where the asterisk (*) here indicates the $\epem$ c.m. frame.
At least one track is required to have a muon identification likelihood ratio $\mathcal{L}(\mu)/(\mathcal{L}(e)+\mathcal{L}(\mu)+\mathcal{L}(K)+\mathcal{L}(\pi)+\mathcal{L}(p)+\mathcal{L}(d))>0.9$, corresponding to a selection efficiency of about 95\% and a $\pi-\mu$ misidentification rate around 5\%.
Here, the identification likelihood $\mathcal{L}$ for each charged particle hypothesis (electron, muon, kaon, pion, proton, and deutron) combines particle-identification information from all subdetectors.
We combine the $\Upsilon(nS)$ candidate with a pair of oppositely charged particles assumed to be pions, individually requiring a minimum transverse momentum of $p_{\rm T}>60$ MeV/$c$ in the laboratory frame and a muon identification likelihood ratio smaller than $0.75$ to suppress the background from muon misidentification. 
To remove potential backgrounds from photon conversions in the detector material, we require the opening angle between the pion candidates to satisfy thre requirement $\cos{\theta_{\pi^{+}\pi^{-}}}<0.9$, where $\theta$ is calculated in the lab frame.

The $\mumu$ pair, $\pipi$ pair, and the overall $\pipi\mumu$ candidate are individually fitted and constrained to originate from the same vertex.
We reject poorly reconstructed events failing the vertex fit.
To suppress the background from events with additional particles, we require the momentum of the signal candidates in the $\epem$ c.m.\ frame $p^{*}(\pi^{+}\pi^{-}\mu^{+}\mu^{-})$ to be less than $100$ MeV/$c$.

To extract signal yields, we define the difference of the overall candidate invariant mass and the $\Upsilon(nS)$ candidate invariant mass as $\Delta M=M(\pi^{+}\pi^{-}\mu^{+}\mu^{-})-M(\mu^{+}\mu^{-})$, which helps to improve the mass resolution.
The selection criteria are optimized with the figure-of-merit $\epsilon_{P} / (a+\sqrt{N_B}), a = 5/2$ \cite{Punzi:2003bu}, using the simulated signal efficiency ($\epsilon_{P}$) and the number of background events ($N_B$) determined from data sidebands to avoid introducing bias.
The sidebands are defined as the region more than 10 standard deviations ($\sigma$) in resolution of the $\Upsilon(10753)$ signal away from the expected signal region in the $\Delta M$ dimension.
The optimization of selection criteria is achieved by progressively refining the selection parameters, at each iteration varying a single parameter while keeping all others constant.
The final values are rounded to the closest whole numbers to ensure consistency among different samples and to avoid over-tuning the optimization.
According to simulation, very few background events are expected from low multiplicity processes after the final selection and they do not produce signal-like structures in the data.

\section{Signal determination}

\begin{figure*}[htb]
    \centering
    \includegraphics[width=0.45\linewidth]{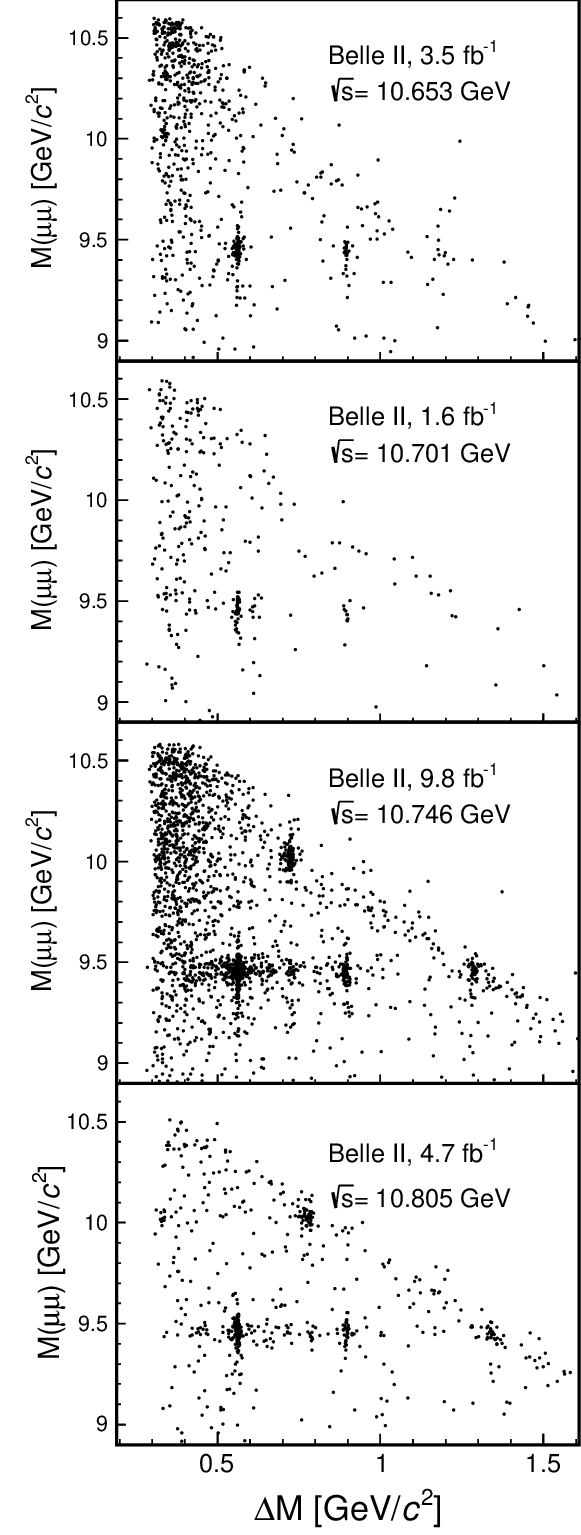}
    \includegraphics[width=0.45\linewidth]{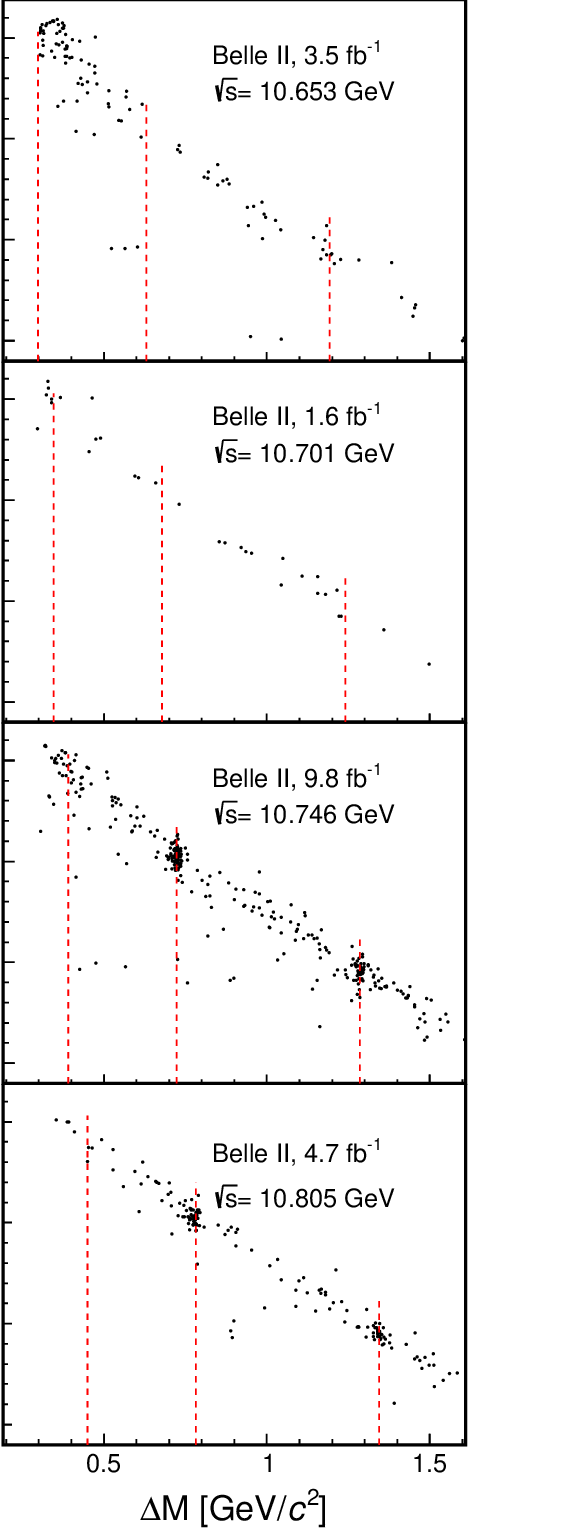}
  \caption{Distribution of dimuon mass as a function of the mass difference between $\pipi\mu^+\mu^-$ and $\mu^+\mu^-$ system for (left) all events and (right) for events with c.m.~momentum of the $\pipi\mu^+\mu^-$ system smaller than 100 MeV/c. Dashed lines indicate the positions of the $\Upsilon(10753)\to\pipi\Upsilon(nS)$ signals.}
  \label{fig:2Dplot}
\end{figure*}

Figure~\ref{fig:2Dplot} shows two-dimensional distributions of the dimuon invariant mass $M(\mu^{+}\mu^{-})$ as a function of the mass difference $\Delta M$ for data taken at each collision energy with and without the $p^{*}(\pi^{+}\pi^{-}\mu^{+}\mu^{-})<100~\rm MeV$ momentum requirement. 
For each $\Upsilon(nS)$ state, we define our analysis region by requiring
$(\Delta M_{\rm nom} - 100) < \Delta M < (\Delta M_{\rm nom} +70)~\mevcc$. 
The expected peak position $\Delta M_{\rm nom}$ is $\sqrt{s} - m_{\Upsilon(nS)}$, where $m_{\Upsilon(nS)}$ is the known $\Upsilon(nS)$ mass~\cite{pdg}: the three positions are shown as the vertical dashed lines in the right panels of Fig.~\ref{fig:2Dplot}.
The width of these analysis regions are ten times the signal resolution ( $\pm10\sigma$), and are used as the fitting ranges in signal yield determination.
Here the resolution is about $10(7)~{\rm MeV}/c^2$ for the lower-(higher-)$\Delta M$ side of the signal.
The signal region is defined as $(\Delta M_{\rm nom} - 30) < \Delta M < (\Delta M_{\rm nom} + 21)~\mevcc$.

In addition to clusters of events in the signal regions, concentrations of events from ISR and cascade processes are clearly visible in the left column in Fig.~\ref{fig:2Dplot}.
For example, $\epem\to\gamma_{\rm ISR}\Upsilon(2S),~\Upsilon(2S)\to\pipi\Upsilon(1S)$, corresponding to the largest event cluster on the very left in each panel.
Before any consideration of the signal regions, we first measure the cross sections for the $\epem\to\gamma_{\rm ISR}\Upsilon(3S,2S)$ control
processes in the full energy-scan datasets as well as the process $\Upsilon(4S)\to\pipi\Upsilon(1S,2S)$. This offers a thorough consistency-check of the signal-extraction approach and of dimuon reconstruction efficiency. The results are consistent with expectations~\cite{pdg}.

\begin{figure}[htb]
\flushleft
    \includegraphics[width=0.32\linewidth]{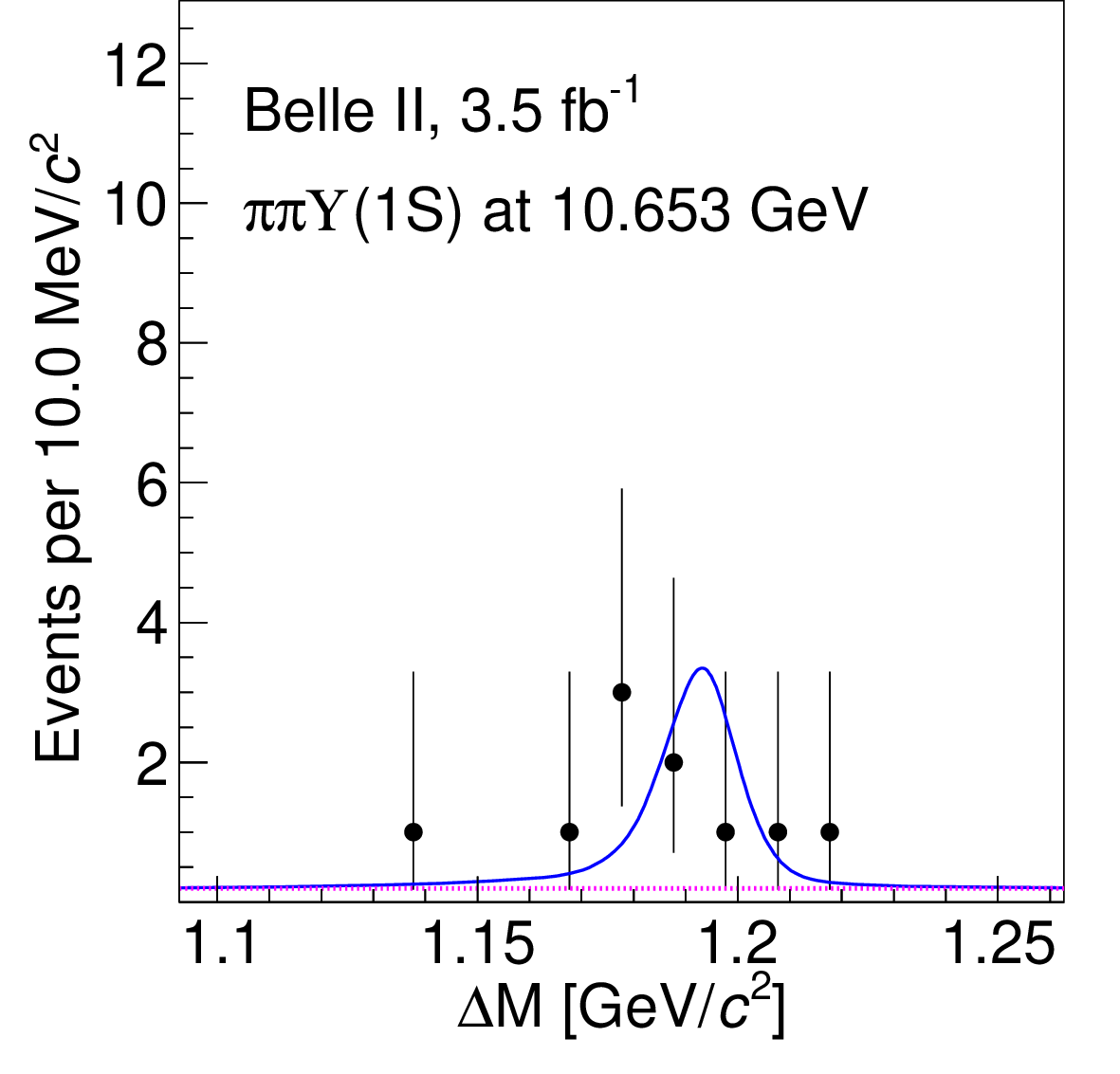}
    \includegraphics[width=0.32\linewidth]{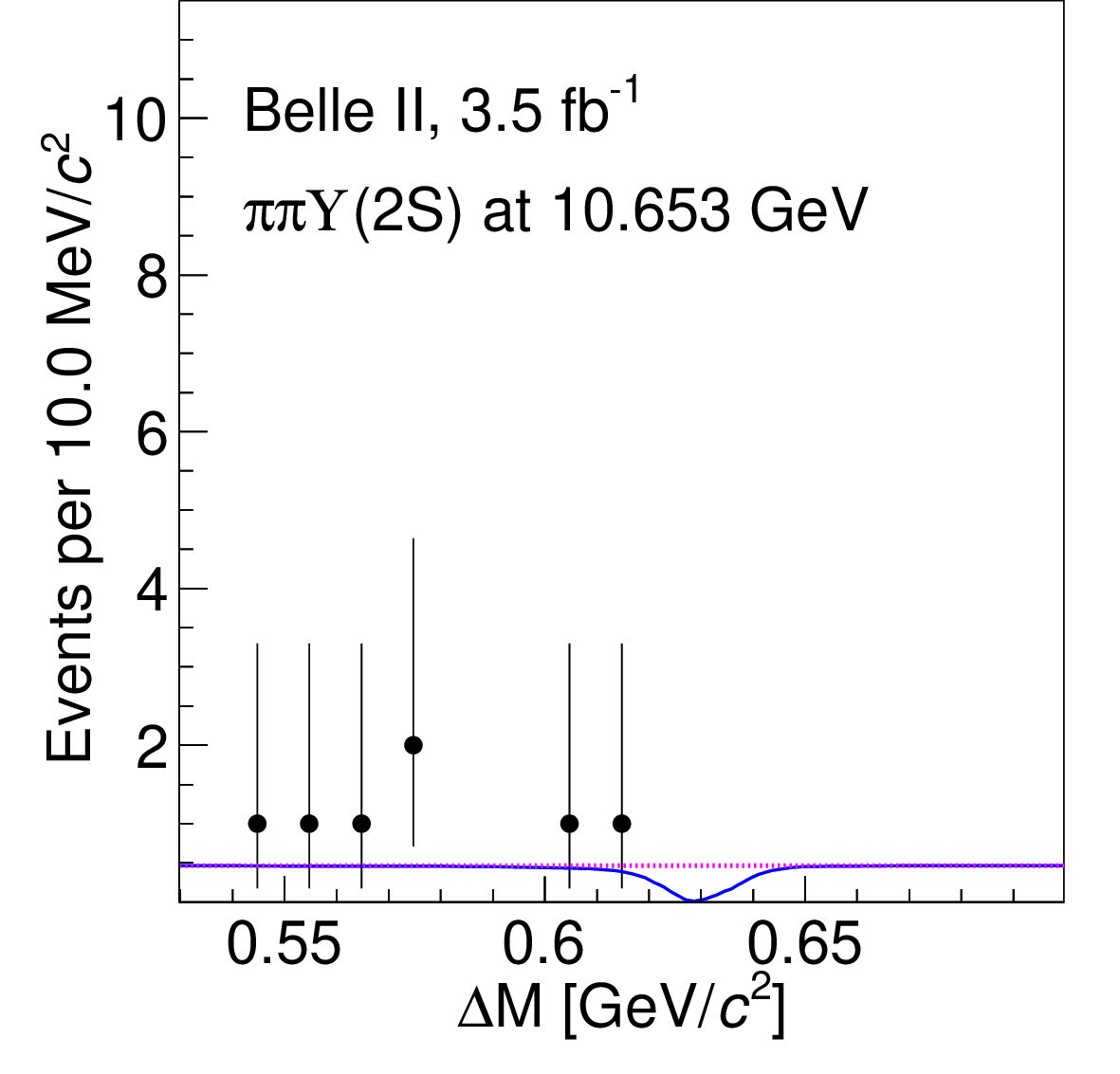}\\
    \includegraphics[width=0.32\linewidth]{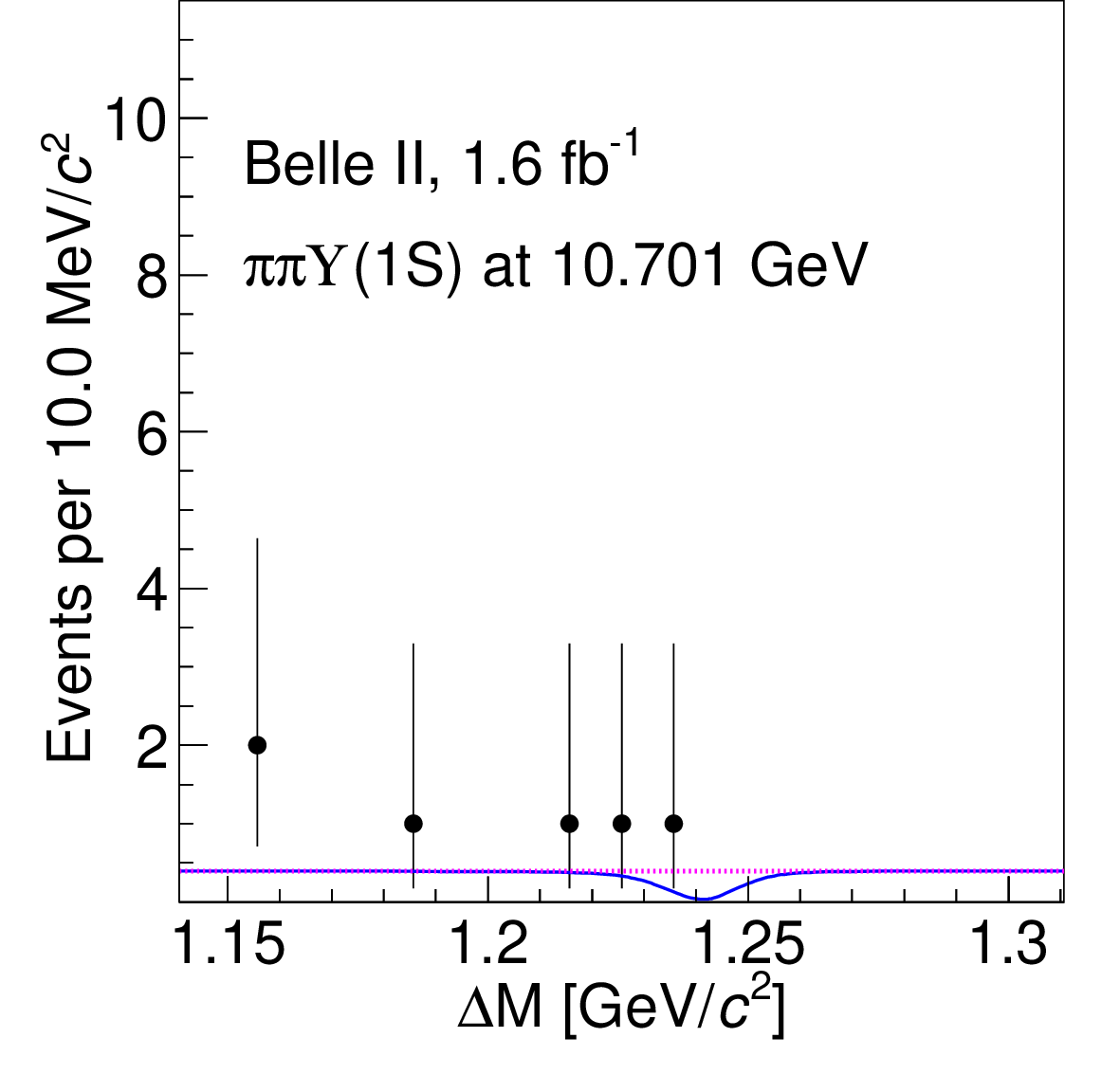} 
    \includegraphics[width=0.32\linewidth]{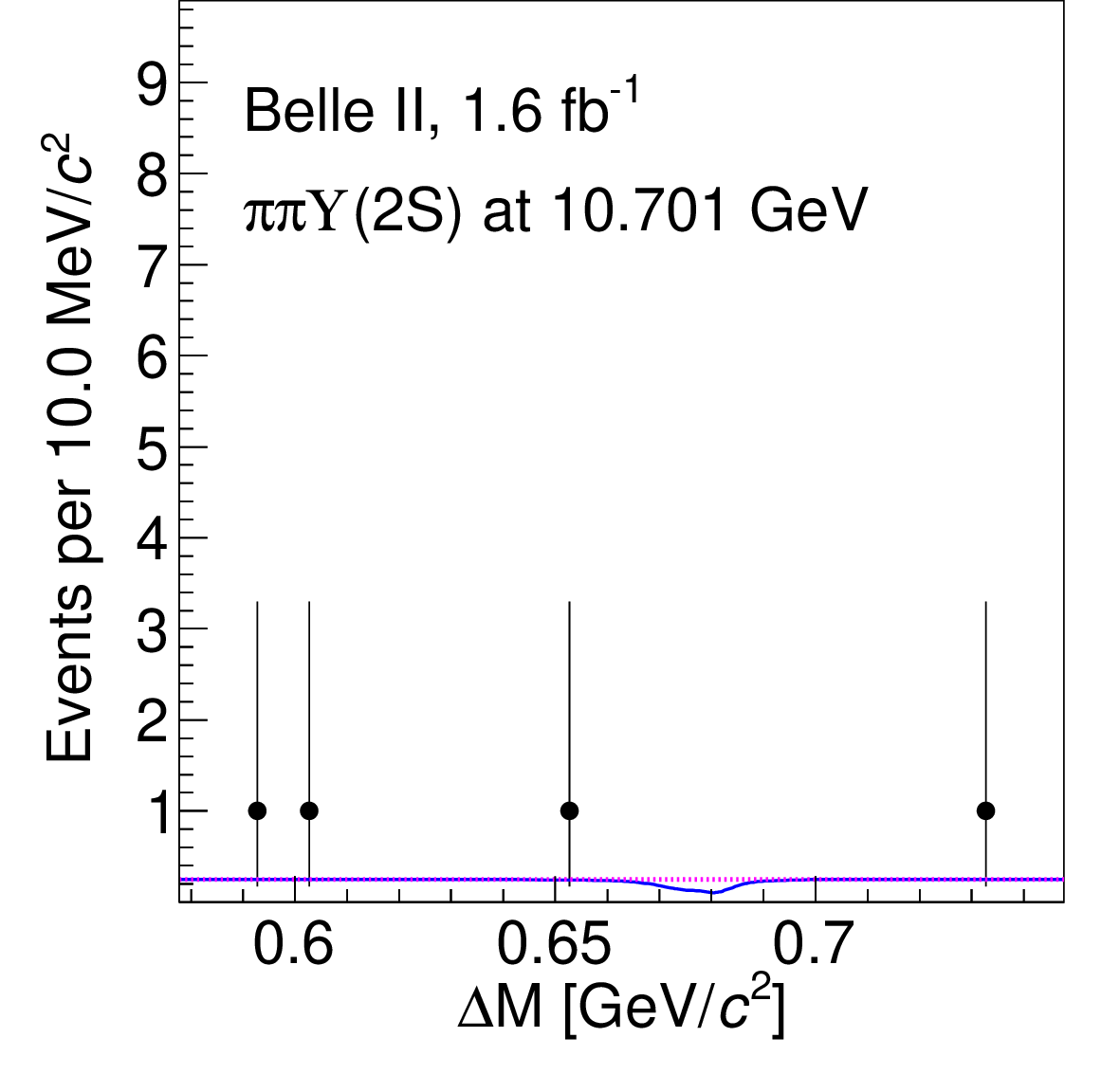}
    \includegraphics[width=0.32\linewidth]{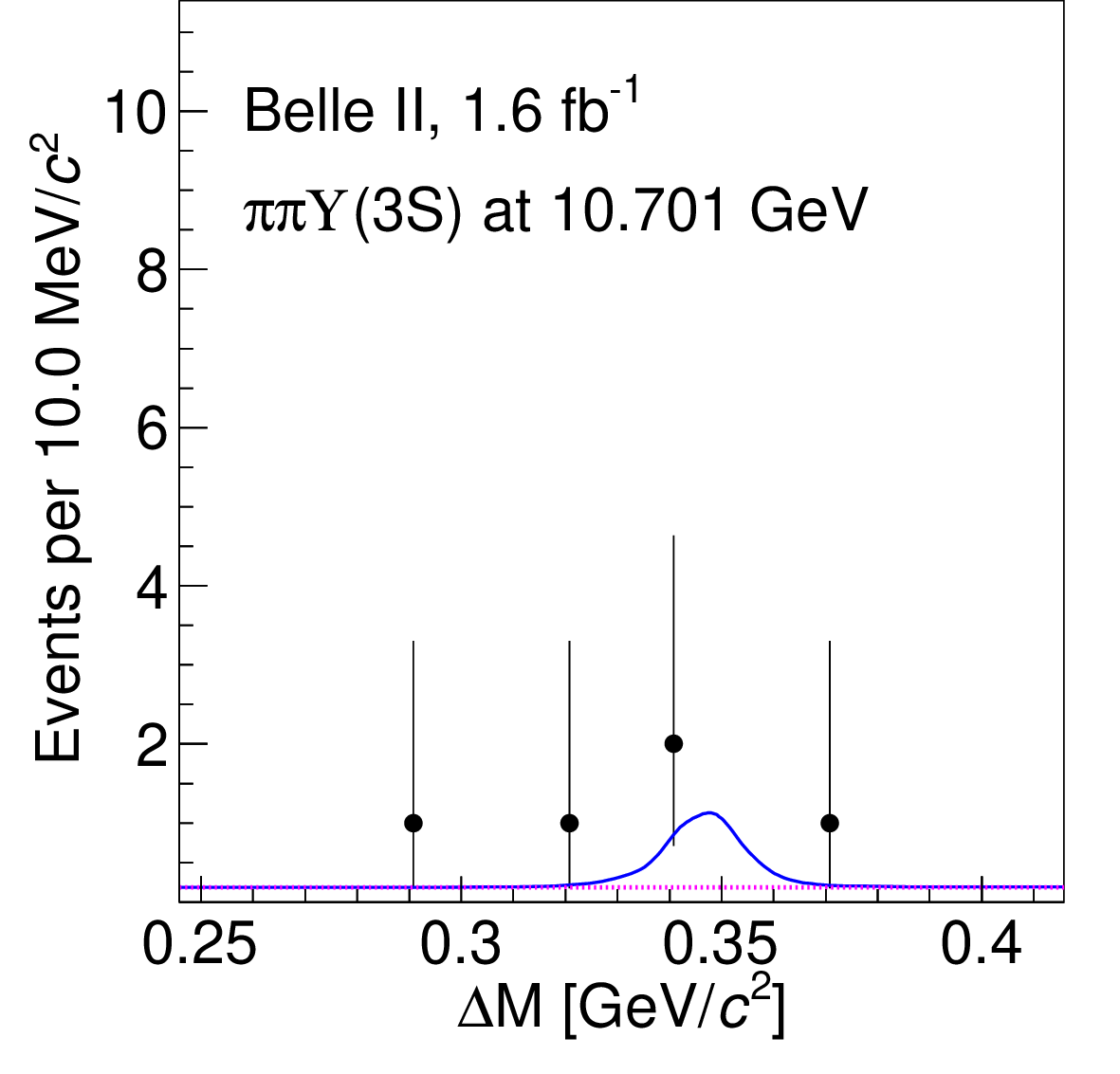}
    \includegraphics[width=0.32\linewidth]{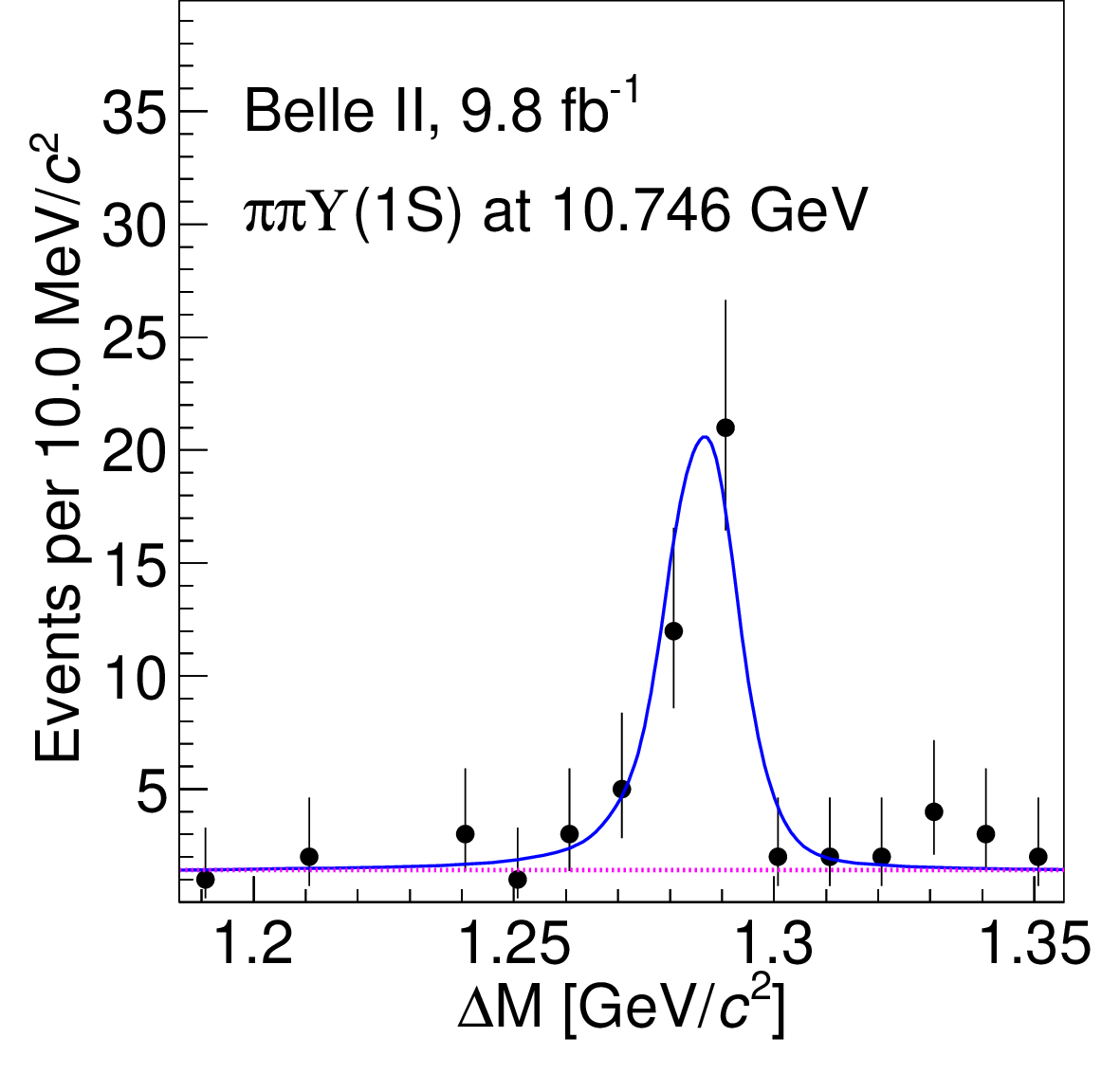}
    \includegraphics[width=0.32\linewidth]{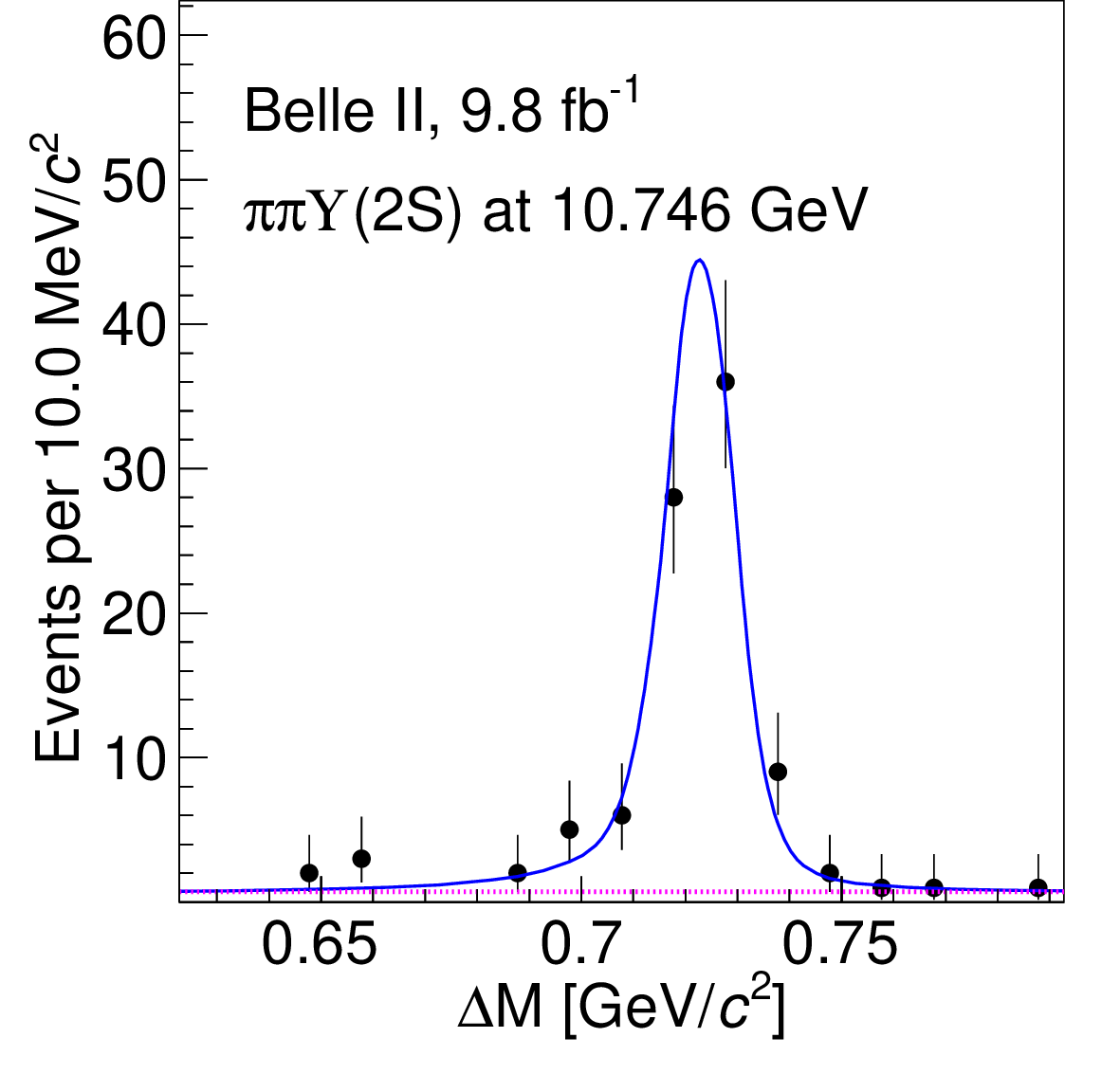}
    \includegraphics[width=0.32\linewidth]{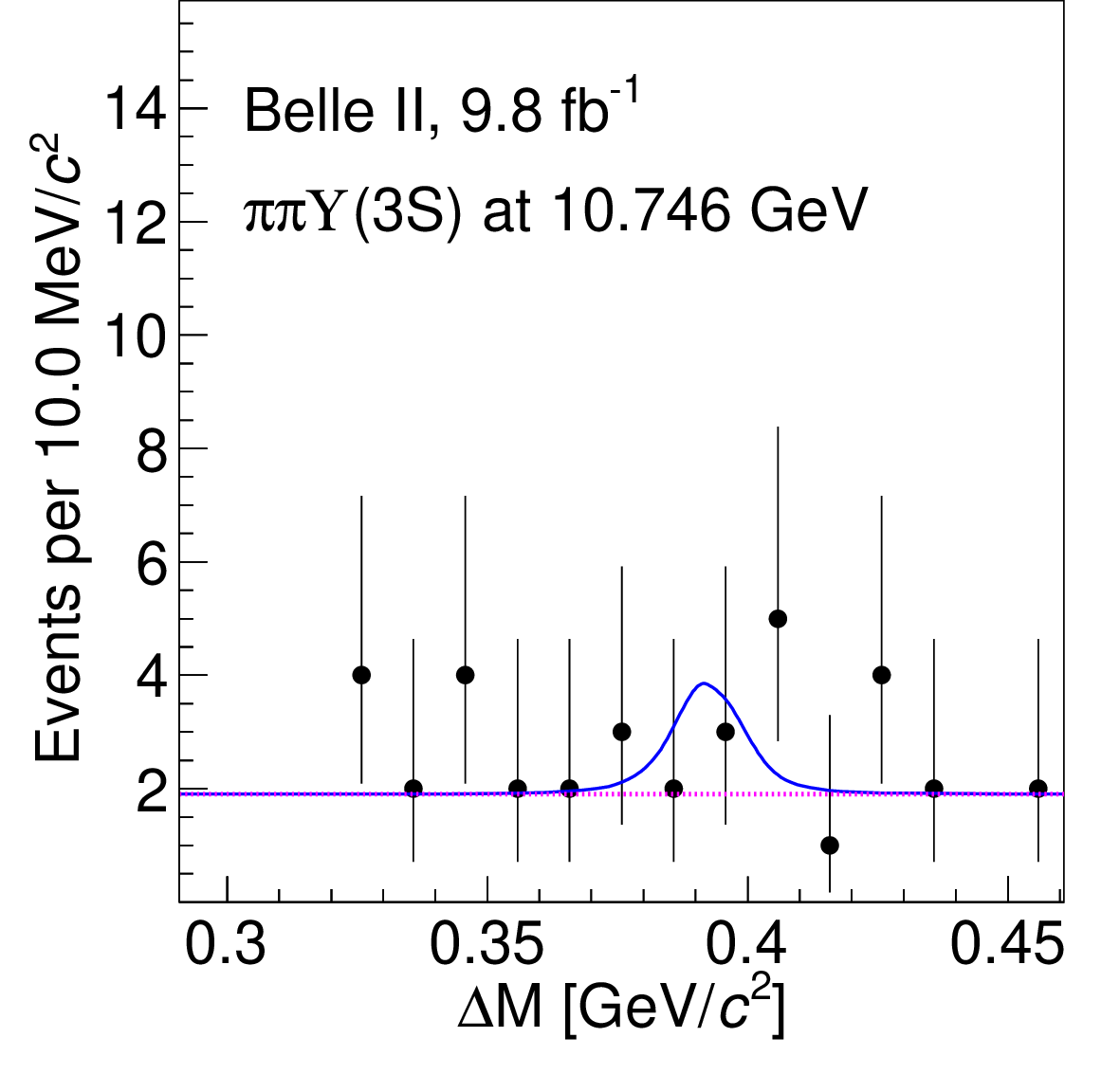}
    \includegraphics[width=0.32\linewidth]{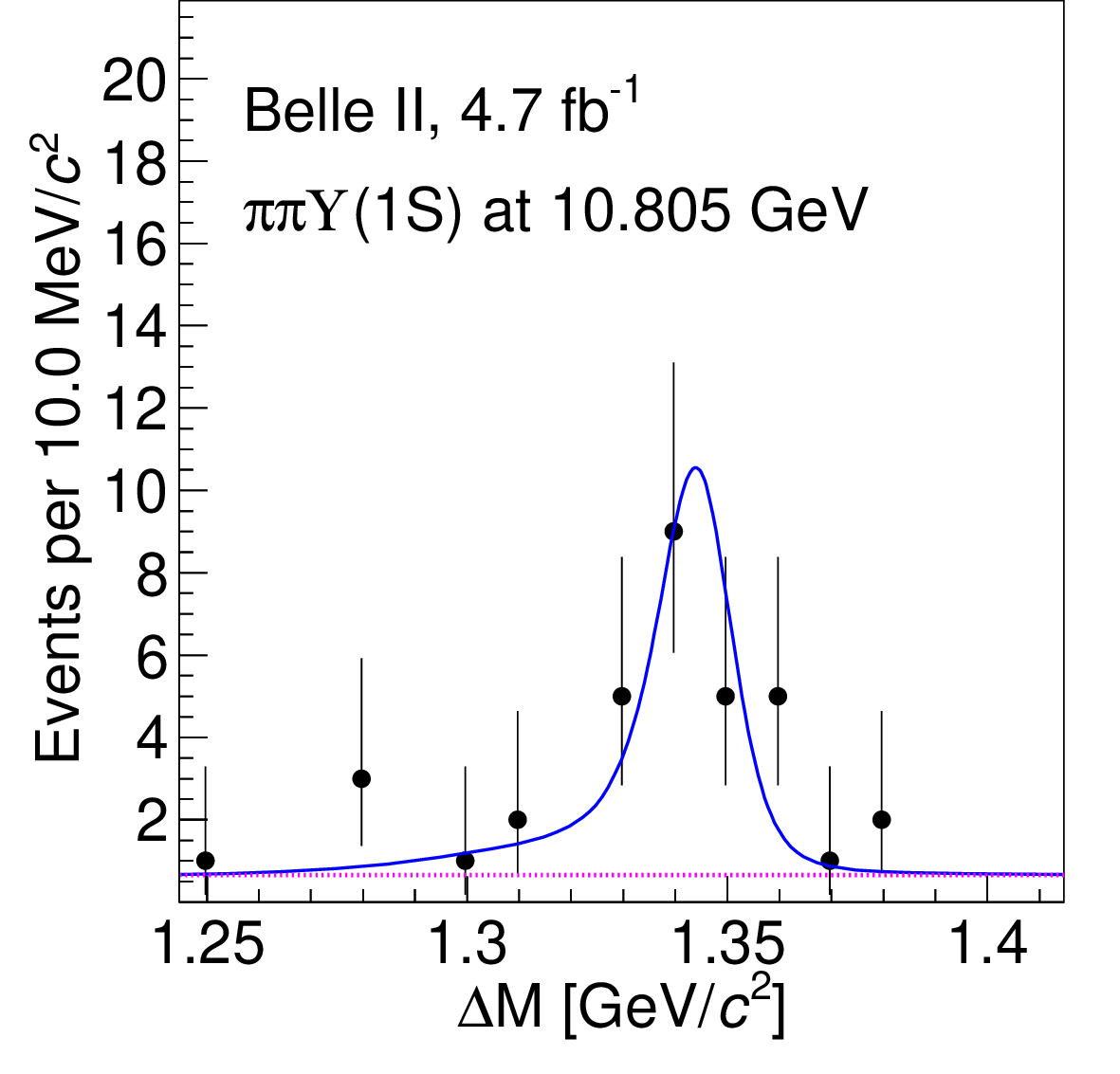}
    \includegraphics[width=0.32\linewidth]{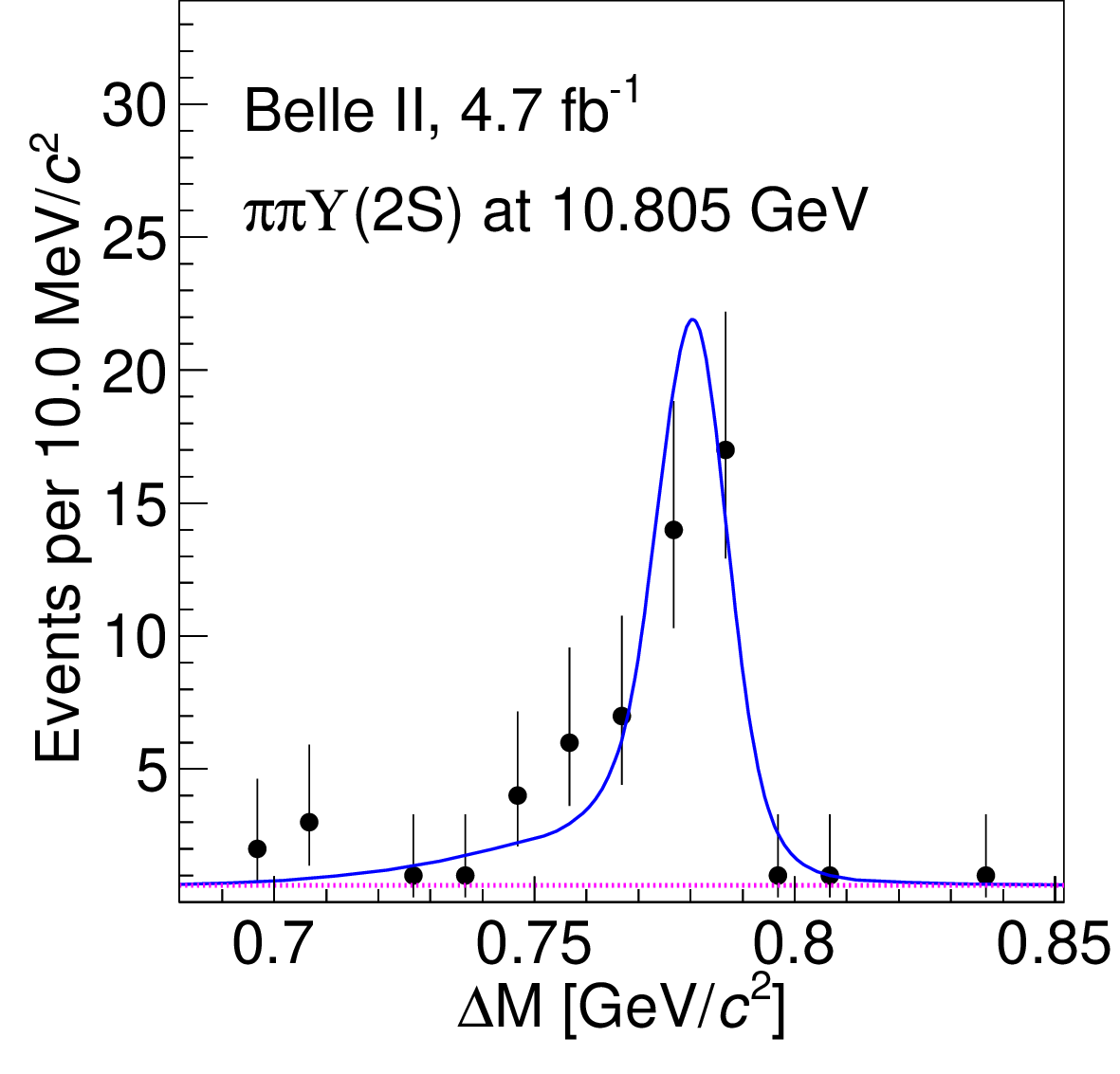}
    \includegraphics[width=0.32\linewidth]{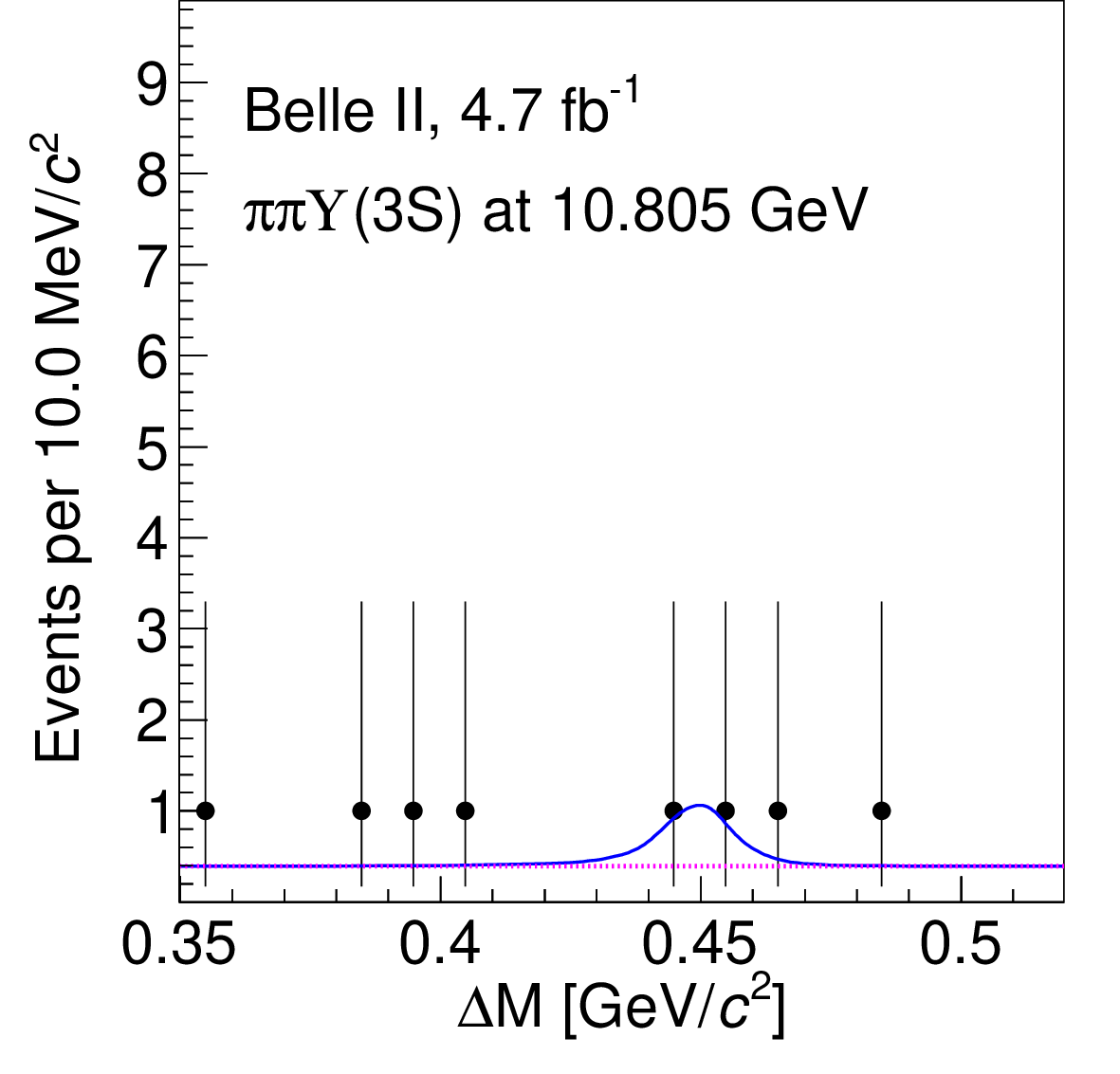}
  \caption{Distributions of the difference between the $\pipi\mumu$ mass and the dimuon mass with fit results overlaid. The points with error bars are the data, the solid blue curve is the fit result, and the dashed magenta curve is the background component. 
  The plots from left to right refer to $\pi^{+}\pi^{-}\Upsilon(1S)$, $\pi^{+}\pi^{-}\Upsilon(2S)$, and $\pi^{+}\pi^{-}\Upsilon(3S)$ candidates and from top to bottom correspond to data taken at energies $\sqrt{s}=$ 10.653, 10.701, 10.746, and 10.805 GeV, respectively. }
  \label{fig:fit}
\end{figure}

The projections of the $\Delta M$ distributions in the analysis regions for events restricted to the $p^{*}(\pi^{+}\pi^{-}\mu^{+}\mu^{-})$ signal region are shown in Fig.~\ref{fig:fit} for all energy points and analysis channels.
Signal for $\Upsilon(1S)$ and $\Upsilon(2S)$ resonant structures are found for $\epem\to\pipi\Upsilon(1S,2S)$ in datasets with $\sqrt{s}=10.746$ and $10.805$ GeV.
Maximum likelihood fits to the unbinned $\Delta M$ distributions in each channel are used to obtain the signal yields. 
Two components are included in the fit: signal and background.
The probability density functions of the signal distributions are derived from simulation and are then weighted according to the result of the amplitude and the resulting Born cross-section lineshape, as explained in detail in Sec. 5.
The background in this limited fit range is well-described by a linear function.

\begin{figure}[!h]
    \centering
    \includegraphics[width=0.45\linewidth]{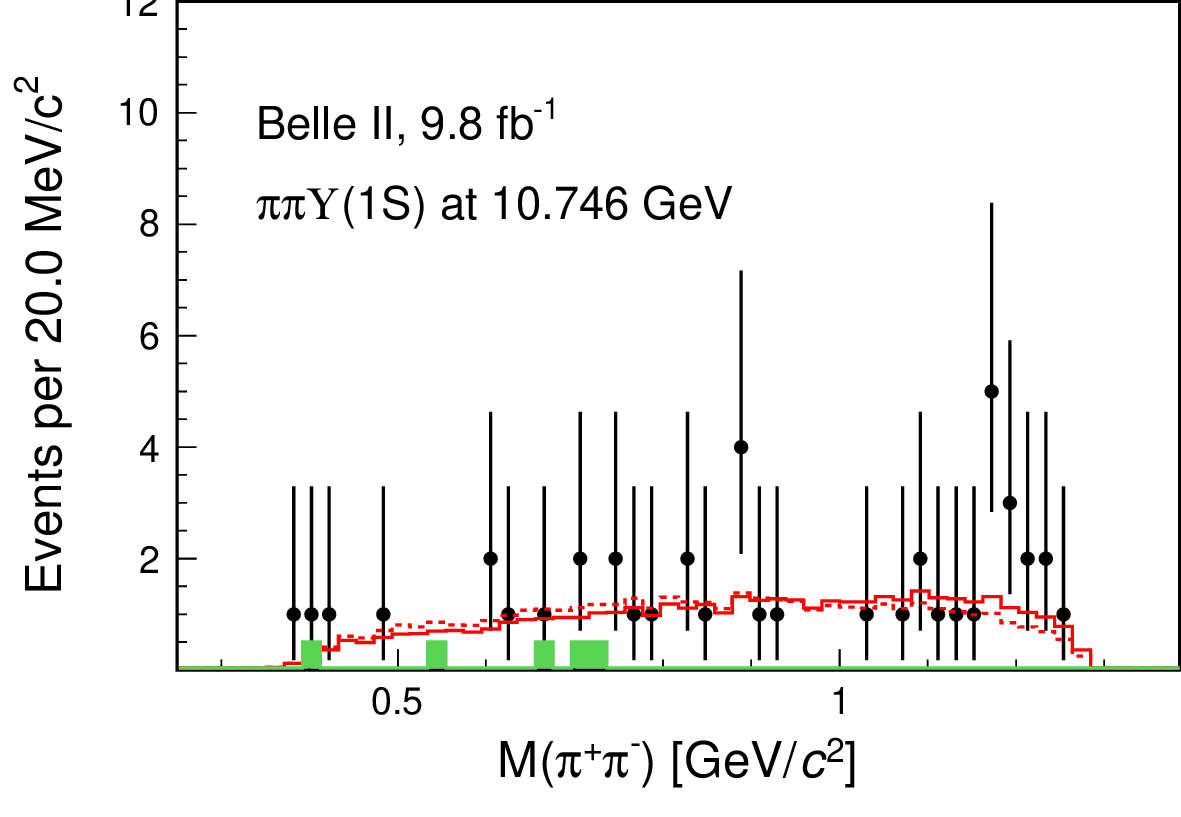}
    \includegraphics[width=0.45\linewidth]{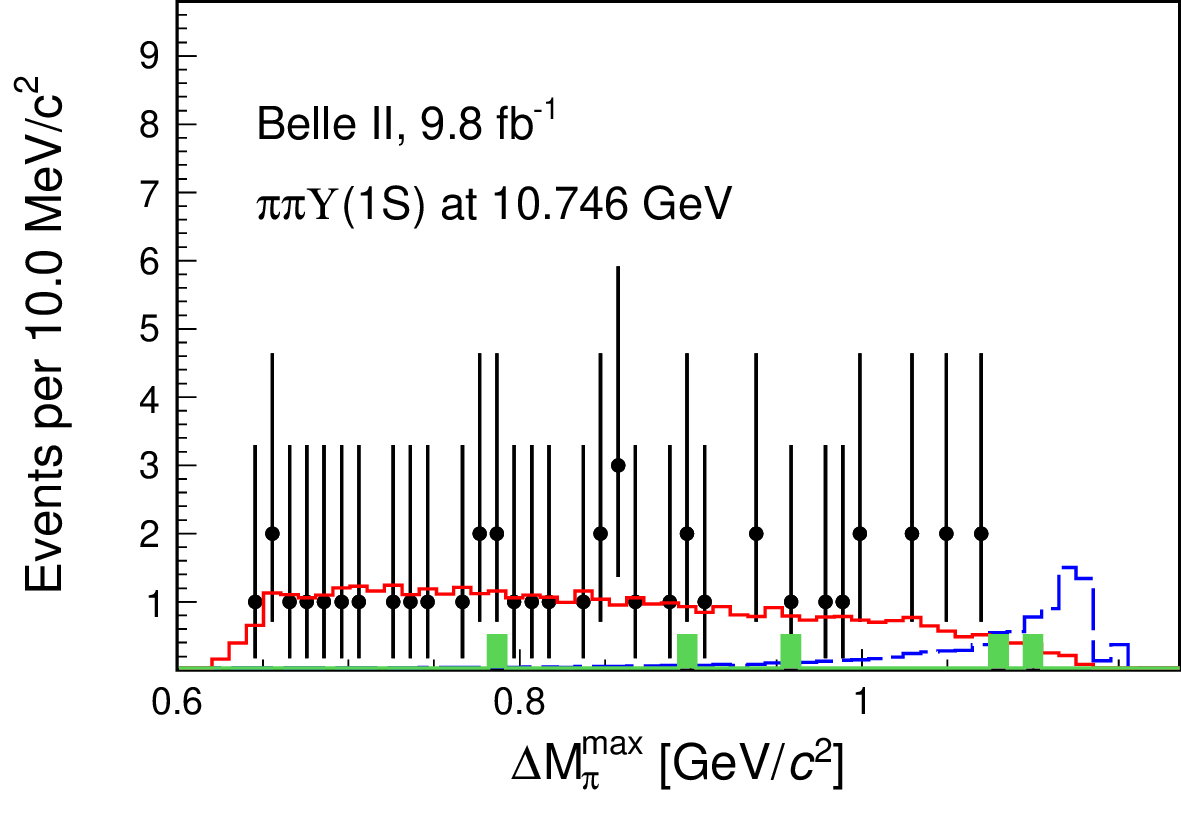}
    \includegraphics[width=0.45\linewidth]{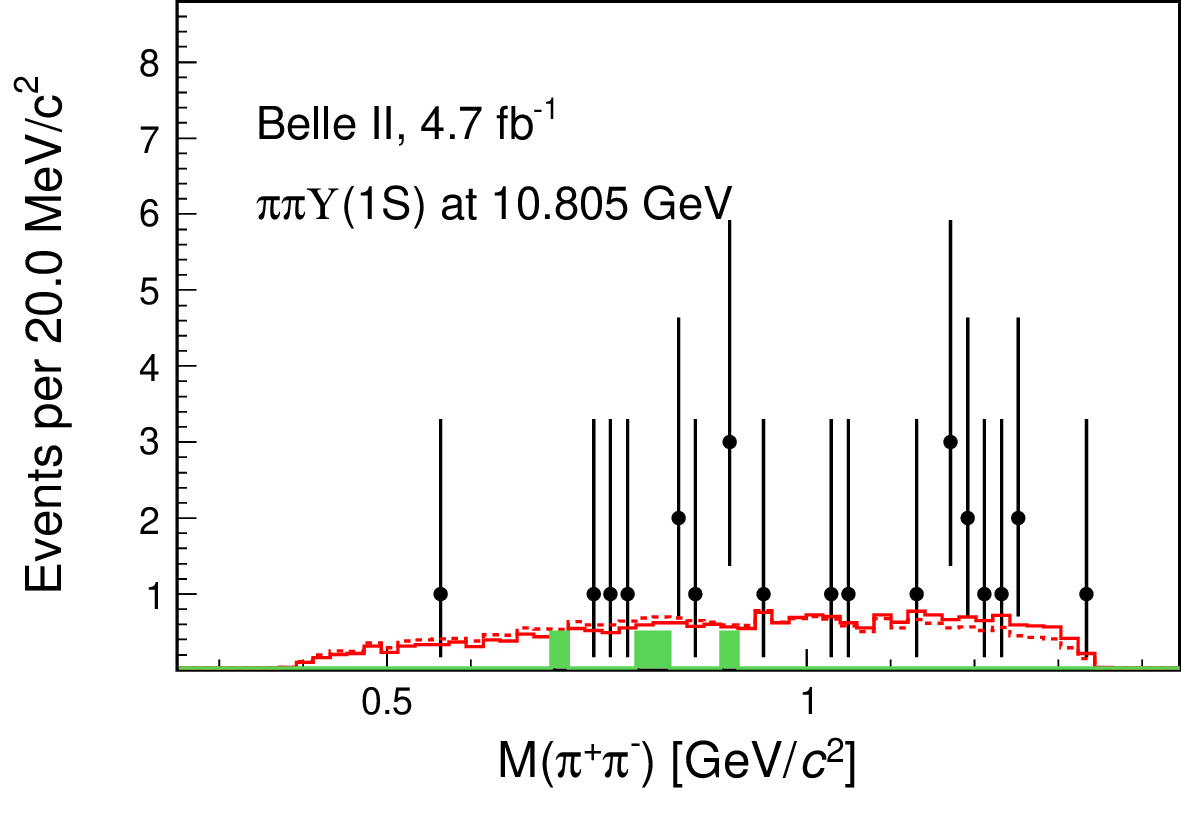}
    \includegraphics[width=0.45\linewidth]{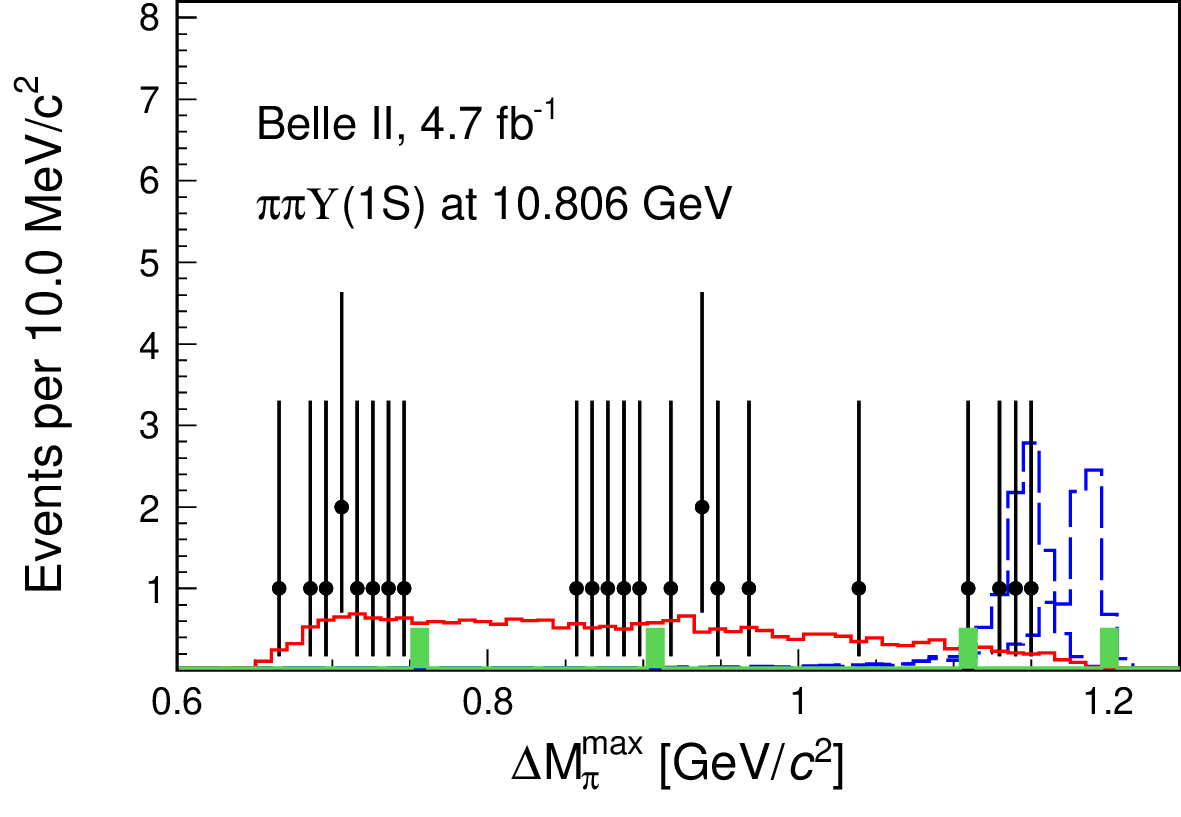}
    \includegraphics[width=0.45\linewidth]{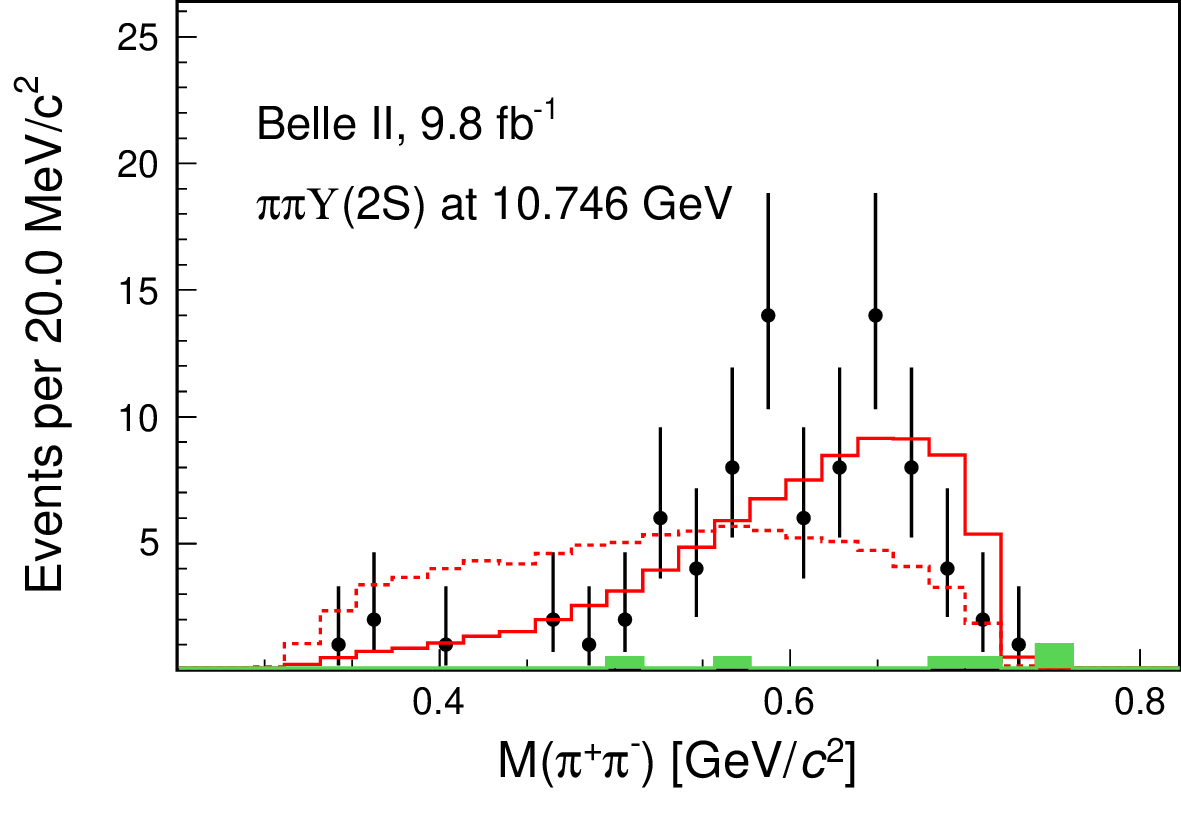}
    \includegraphics[width=0.45\linewidth]{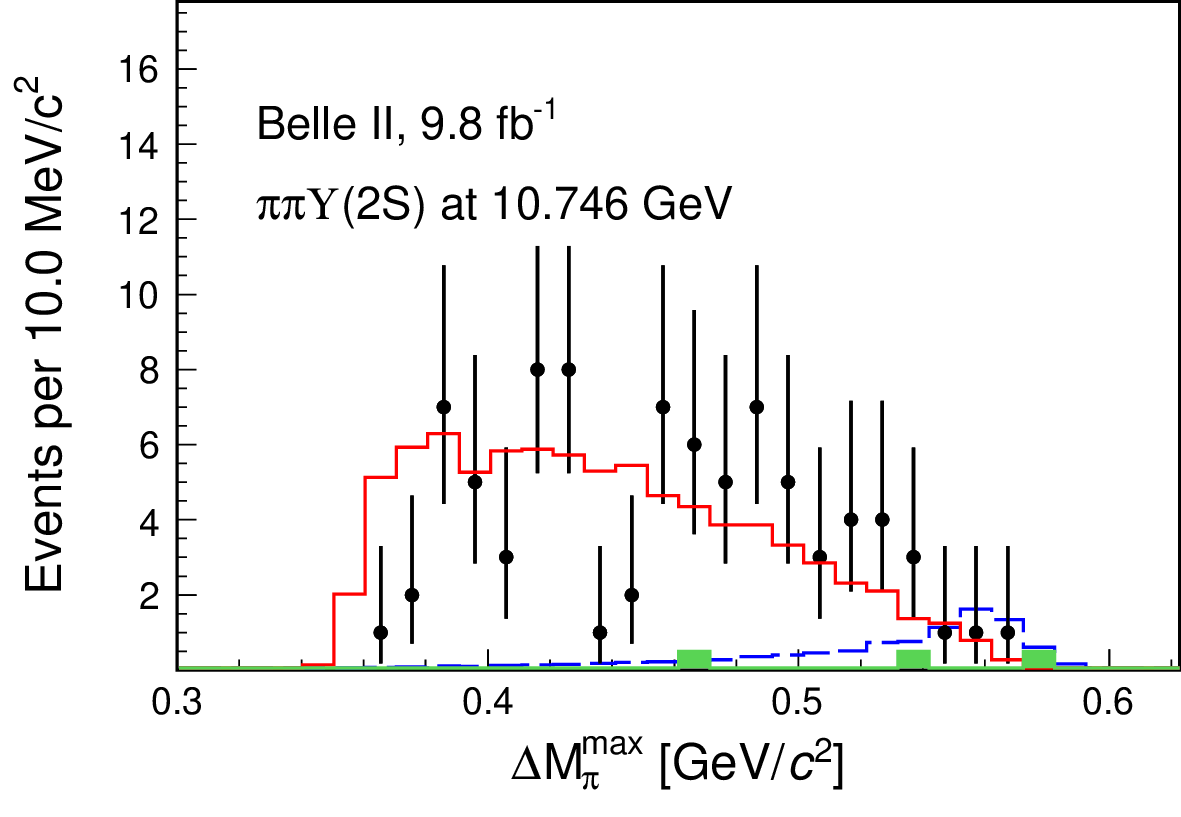}
    \includegraphics[width=0.45\linewidth]{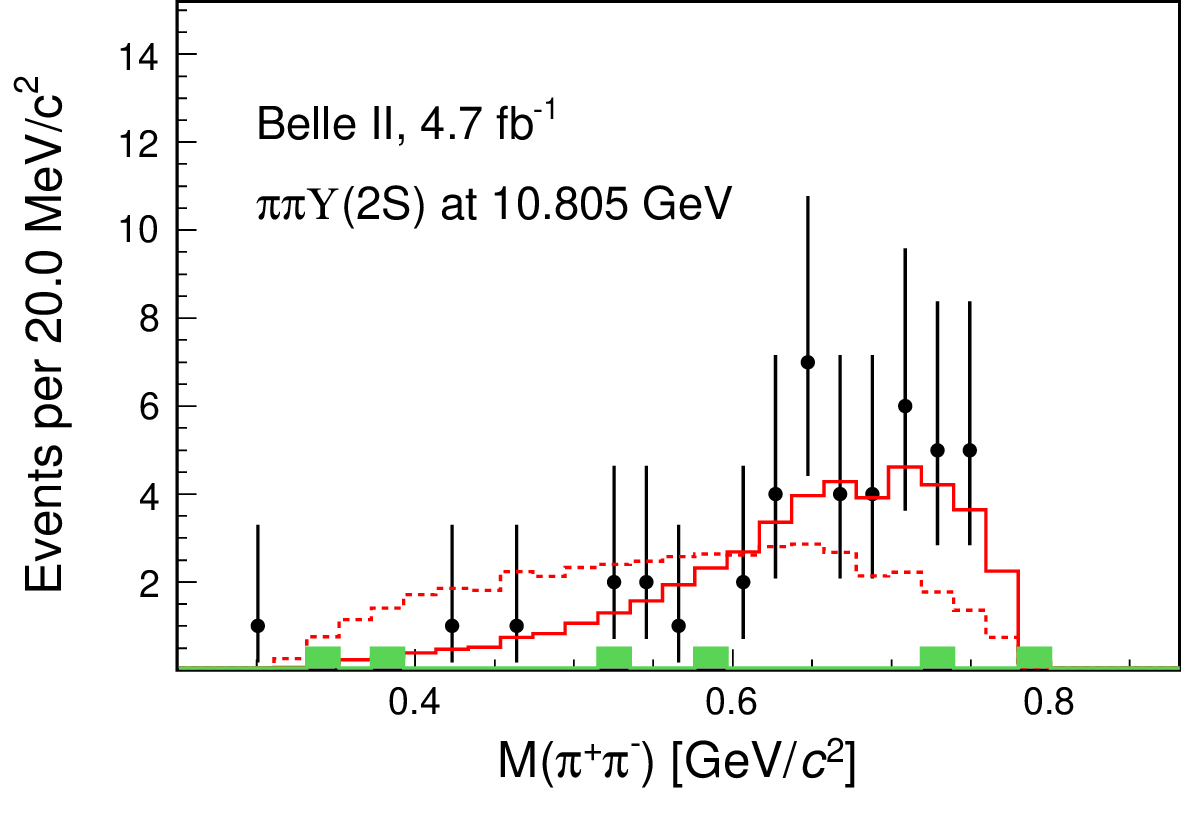} 
    \includegraphics[width=0.45\linewidth]{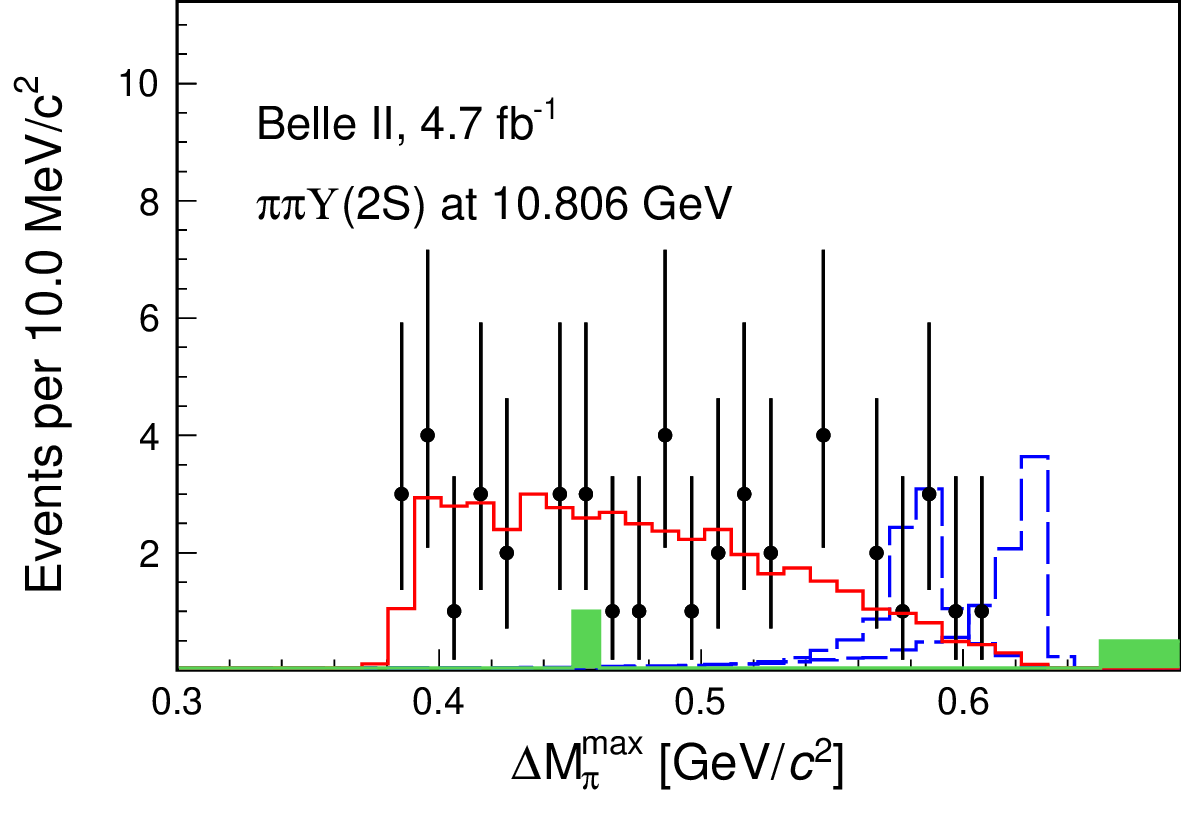}
  \caption{Distributions of dipion mass (left) and maximal difference between the $\pi^{\pm}\mumu$ mass and the $\mumu$ mass (right). Plots from top to bottom show $\pi^{+}\pi^{-}\Upsilon(1S)$ at $\sqrt{s}=10.746$ GeV, $\pi^{+}\pi^{-}\Upsilon(1S)$ at $\sqrt{s}=10.805$ GeV, $\pi^{+}\pi^{-}\Upsilon(2S)$ at $\sqrt{s}=10.746$ GeV, and $\pi^{+}\pi^{-}\Upsilon(2S)$ at $\sqrt{s}=10.805$ GeV. Points with error bars show the events in the signal region from data, green shaded histograms show the events in the sideband region, red histograms are the weighted simulated signal, red dashed histograms are the phase space signal simulation, and blue dashed histograms are the $Z_b(10610/10650)^{\pm}$ from simulation. The simulated signal sample is normalized to the number of events in data, while simulated $Z_b(10610/10650)^{\pm}$ events are normalized arbitrarily.}
  \label{fig:mpipi}
\end{figure}

Since Belle has previously observed a $Z_b(10610/10650)^\pm\to\pi^{\pm}\Upsilon(nS)$ signal in $\Upsilon(5S)\to\pipi\Upsilon(nS)$ decays~\cite{Belle:2011aa}, 
we search for intermediate decays in $\pi^{+}\pi^{-}\Upsilon(1S)$ and $\pi^{+}\pi^{-}\Upsilon(2S)$ signal candidates at $\sqrt{s}$ = 10.746 and 10.805 \gev, where significant $\Upsilon(10753)$ signals are observed.
We examine the dipion invariant mass, $M(\pi^{+}\pi^{-})$, and the mass difference between the invariant mass of the $Z_b^\pm$ and $\Upsilon(nS)$ candidate, $\Delta M_{\pi} = M(\pi^\pm\mu^+\mu^-)-M(\mu^+\mu^-)$.
The $\Delta M_{\pi}$ variable provides improved resolution for reconstructing $Z_b^\pm$ candidates.
Because there are two pions per event, we choose the larger value of $\Delta M_{\pi}$ as the $Z_{b}^\pm$ candidate.
The distributions for events in the signal regions are shown in Fig. \ref{fig:mpipi} for $M(\pi^{+}\pi^{-})$ (left) and $\Delta M_{\pi}$ (right), compared with the events from sideband regions.
Since the sideband contributions are small, we neglect them in the studies that follow.
No signals for $Z_b(10610)^{\pm}$ or $Z_b(10650)^{\pm}$ are evident, and the simulated phase-space distribution is consistent with the data.

The dipion invariant mass in $\pi^{+}\pi^{-}\Upsilon(1S)$ is also consistent with the simulated phase-space distribution. 
This is not the case for $\pi^{+}\pi^{-}\Upsilon(2S)$, in which the dipion mass is similar to that of, \emph{e.g.}, $\Upsilon(2S)\to\pipi\Upsilon(1S)$~\cite{CLEO:2007rbi}.
To better represent the data, we perform a fit to determine the amplitude, and weight the phase space simulation accordingly.
An extended unbinned maximum likelihood fit is performed to the four-momenta of the $\pi^+$ and $\pi^-$ from the samples at $\sqrt{s}=10.746$ and $10.805$ GeV.
We use the following formula to describe the amplitude~\cite{CLEO:2007rbi},
\begin{equation}
\mathcal{M}\propto \mathcal{A}(q^2-2m^2_{\pi}) + \mathcal{B}E_1E_2+\mathcal{C}[(\lambda' q_1)(\lambda q_2)+(\lambda' q_2)(\lambda q_1)], 
\label{amp_y2s}
\end{equation}
where $q^2$ is the invariant mass squared of the pion pair, $m_{\pi}$ is the pion mass~\cite{pdg}, $q_{i}$ and $E_{i}$ are the four-momenta and energy of the $i$th pion in the $\Upsilon(10753)$ rest frame, respectively, $\lambda^{(')}$ is the polarization vector of the parent (final-state) $\Upsilon$, and $\mathcal{A}$, $\mathcal{B}$ are unconstrained complex parameters.
The $\mathcal{C}$ term couples the transitions via the chromo-magnetic moment of the bound-state $b$ quark, and hence requires a spin flip.
This term is expected to be highly suppressed by the large mass of the $b$ quark, as confirmed by CLEO findings~\cite{CLEO:2007rbi}; hence, we assume $\mathcal{C}$ to vanish.
In the fit, we fix the real and imaginary parts of the parameter $\mathcal{A}$ to be one and zero, respectively, allow $\mathcal{B}$ to vary, and fix $\mathcal{C}$ to be zero.
Here fixing $\mathcal{A}$ to the arbitrary (1,0) values is allowed as only relative differences with regard to $\mathcal{B}$ have physical significance.
From the fit, we obtain $1.1 \pm 0.3$ and $4.7 \pm 1.5$ for the real and imaginary parts of $\mathcal{B}$.
The uncertainties here are statistical only.

The $M(\pi^+\pi^-)$ projection of the fit results is compared with the data for $\pi^{+}\pi^{-}\Upsilon(2S)$ and is qualitatively satisfactory: the reduced $\chi^2$ values are 1.40 and 0.77 for 10.746 and 10.805 GeV, respectively.
We reweight the simulated signal sample distributions based on this amplitude fit result.
Note that this weight has negligible effect on the shape of the projection of $\Delta M_{\pi}$ for signal events.

\section{Cross section measurement and resonance parameters determination}

We use an iterative approach to compute the Born cross sections of the $\epem\to\pipi\Upsilon(nS)$ process as the true lineshape of the Born cross sections is not known \emph{a priori}, and the reconstruction efficiency estimation and signal shapes will be affected by this lineshape.
In the simulation, the default lineshape of the $\epem$ annihilation cross section is assumed to be that of $\epem\to\mumu$.
In the first iteration, we weight the simulated signal samples according to the $\epem\to\pipi\Upsilon(nS)$ lineshape reported in the Belle publication~\cite{exp_belle}.
We use the weighted simulated signal samples to extract the signal shapes and the reconstruction efficiencies.

The number of signal events, $N_{\rm S}$, is determined by performing an extended unbinned maximum-likelihood fit to $\Delta M$ in each analysis region (Fig. \ref{fig:fit}). 
We allow the number of signal events to be negative in the fit.

The Born cross sections are calculated using 
\begin{equation} \label{eq:bxs}
\sigma_{\rm B} = \frac{N_{\rm S}\,|1-\Pi|^2} {{\cal L}\,\varepsilon\, \mathcal{B}\,(1+\delta)},
\end{equation}
where ${\cal L}$ is the integrated luminosity, 
$\varepsilon$ is the reconstruction efficiency weighted by the intermediate $\pipi$ amplitude and the Born cross section event-by-event, 
$\mathcal{B}$ is the branching fraction of $\Upsilon(nS)\to\mumu$, 
$|1-\Pi|^2$ is the vacuum polarization factor~\cite{vp}, 
and $(1+\delta)$ is the radiative correction factor~\cite{Kuraev:1985hb,Benayoun:1999hm} calculated using the Born cross-section lineshape with the following formula,
\begin{equation}
(1+\delta) = \frac{\int_{0}^{x_m} \sigma_{\rm B}(s(1-x))W(x,s)dx}{\sigma_{\rm B}(s)},
\end{equation}
where $\sigma_{\rm B}(s)$ is the energy-dependent Born cross-section lineshape; and the parameter $x$ and the radiator function $W(x,s)$ are as described by Eq. (8) in Ref.~\cite{Benayoun:1999hm}.
The upper limit of the integration is $x_m=1-s_m/s$, where $s_m$ is the minimum invariant mass squared of the final state, \emph{i.e.} $[m(\pi^+)+m(\pi^-)+m(\Upsilon(nS))]^2$.
The numerical results are listed in Tab.~\ref{tab:born_results}. It should be noted that the energy-dependent reconstruction efficiencies are not monotonic because of the differing beam-induced background conditions during the data-taking period, as well as the differing ISR reweighting at each energy point.

\begin{table}[h]
\small
  \begin{center}
    \begin{tabular}{l c c c c c c c}
      \hline Mode & $N_{S}$  & $\mathcal{L}$ ($\rm pb^{-1}$) & $\epsilon$ & $\mathcal{B}$ ($\%$) & $(1+\delta)$ & $|1-\Pi|^{2}$ & $\sigma_{B}$ (pb)\\
        \hline \multicolumn{2}{l}{($10653.30\pm1.14$) MeV} & & & & \\
        \hline
        $\pi\pi \Upsilon(1S)$ & $\phantom{-}5.1^{+2.9}_{-2.8}$~($1.7\sigma$)    &       3521 & 0.339 & 2.48 & 0.926 & 0.929 &   $\phantom{-}0.17^{+0.10}_{-0.10}\pm0.05$ \\
        $\pi\pi \Upsilon(2S)$ & $-1.0^{+0.8}_{-0.7}$~(-)    &       3521 & 0.476 & 1.93 & 0.672 & 0.929 &   $ -0.04^{+0.03}_{-0.03}\pm0.02$ \\
        \hline \multicolumn{2}{l}{($10700.90\pm0.63$) MeV} & & &  &\\
        \hline
        $\pi\pi \Upsilon(1S)$ & $-1.0^{+0.8}_{-0.7}$~(-)    &       1632 & 0.406 & 2.48 & 0.628 & 0.928 &   $ -0.09^{+0.07}_{-0.06}\pm0.16$   \\
        $\pi\pi \Upsilon(2S)$ & $-0.3^{+0.8}_{-0.5}$~(-)     &       1632 & 0.468 & 1.93 & 0.641 & 0.928 &   $ -0.03^{+0.08}_{-0.05}\pm0.01$  \\
        $\pi\pi \Upsilon(3S)$ & $\phantom{-}1.9^{+2.2}_{-1.5}$~($0.9\sigma$)      &       1632 & 0.161 & 2.18 & 0.578 & 0.928 &   $\phantom{-}0.53^{+0.62}_{-0.42}\pm0.54$  \\
        \hline
        \multicolumn{2}{l}{($10746.30\pm0.48$) MeV} & & & & \\
        \hline
        $\pi\pi \Upsilon(1S)$ & $\phantom{-}41.2^{+7.9}_{-7.3}$~($5.8\sigma$)     &       9818 & 0.421 & 2.48 & 0.588 & 0.930 &   $\phantom{-}0.64^{+0.12}_{-0.11}\pm0.03$  \\
        $\pi\pi \Upsilon(2S)$ & $\phantom{-}84.8^{+10.4}_{-10.1}$~($10.0\sigma$)  &       9818 & 0.489 & 1.93 & 0.597 & 0.930 &   $\phantom{-}1.43^{+0.17}_{-0.17}\pm0.17$ \\
        $\pi\pi \Upsilon(3S)$ & $\phantom{-}3.7^{+4.0}_{-3.3}$~($0.8\sigma$)      &       9818 & 0.264 & 2.18 & 0.578 & 0.930 &   $\phantom{-}0.11^{+0.11}_{-0.09}\pm0.08$ \\
        \hline
        \multicolumn{2}{l}{($10804.50\pm0.70$) MeV} & & &  &\\
        \hline
        $\pi\pi \Upsilon(1S)$ & $\phantom{-}20.7^{+6.6}_{-5.5}$~($3.9\sigma$)     &       4690 & 0.445 & 2.48 & 0.784 & 0.931 &   $\phantom{-} 0.48^{+0.15}_{-0.13}\pm0.04$ \\
        $\pi\pi \Upsilon(2S)$ & $\phantom{-}47.4^{+8.3}_{-7.6}$~($6.3\sigma$)     &       4690 & 0.533 & 1.93 & 0.814 & 0.931 &   $\phantom{-}1.12^{+0.20}_{-0.18}\pm0.11$ \\
        $\pi\pi \Upsilon(3S)$ & $\phantom{-}1.3^{+2.2}_{-1.5}$~($0.6\sigma$)      &       4690 & 0.314 & 2.18 & 0.721 & 0.931 &   $\phantom{-} 0.05^{+0.09}_{-0.06}\pm0.01$ \\
      \hline
\end{tabular}
\end{center}
\caption{Summary of the c.m.\ energies, signal yields and their significance, luminosity, weighted efficiencies, branching fractions of the $\Upsilon(nS)$ decays, ISR correction factors, vacuum polarization factors, and calculation of Born cross sections from the fit results to data. The values in brackets are the statistical significances, and the dash ``-'' denotes that the significance is not available. Uncertainty in the signal yields is statistical only. The first uncertainty in the Born cross section is statistical and the second systematic. The uncertainties in the c.m.\ energy shown here are uncorrelated point-to-point; the correlated uncertainty is 0.5\ MeV.}
\label{tab:born_results}
\end{table}

\begin{figure}[!h]
\flushleft
    \includegraphics[width=0.95\linewidth]{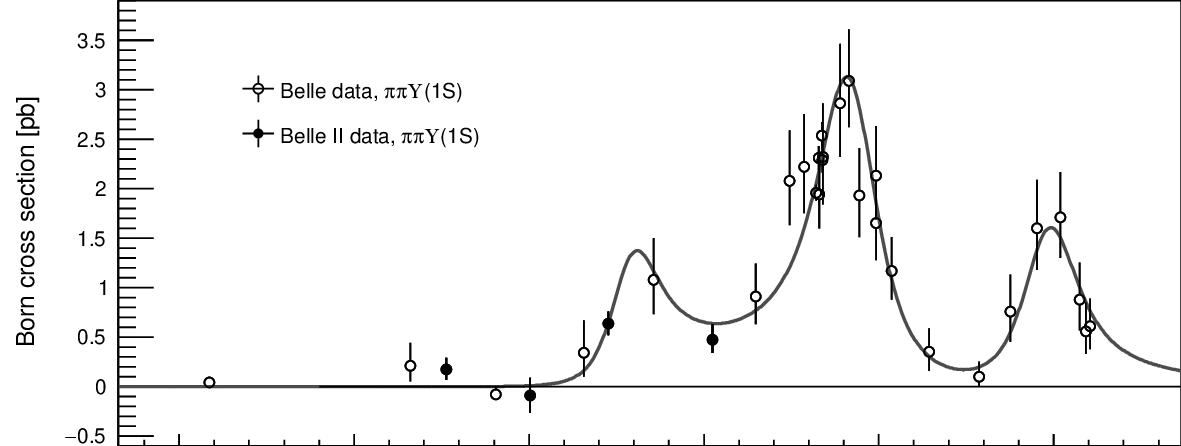}    
    \includegraphics[width=0.95\linewidth]{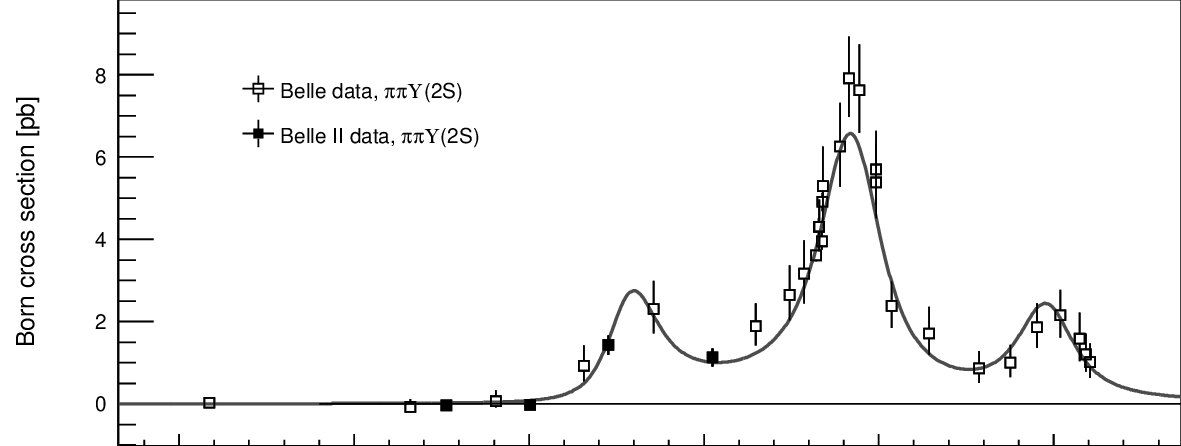}    
    \includegraphics[width=0.95\linewidth]{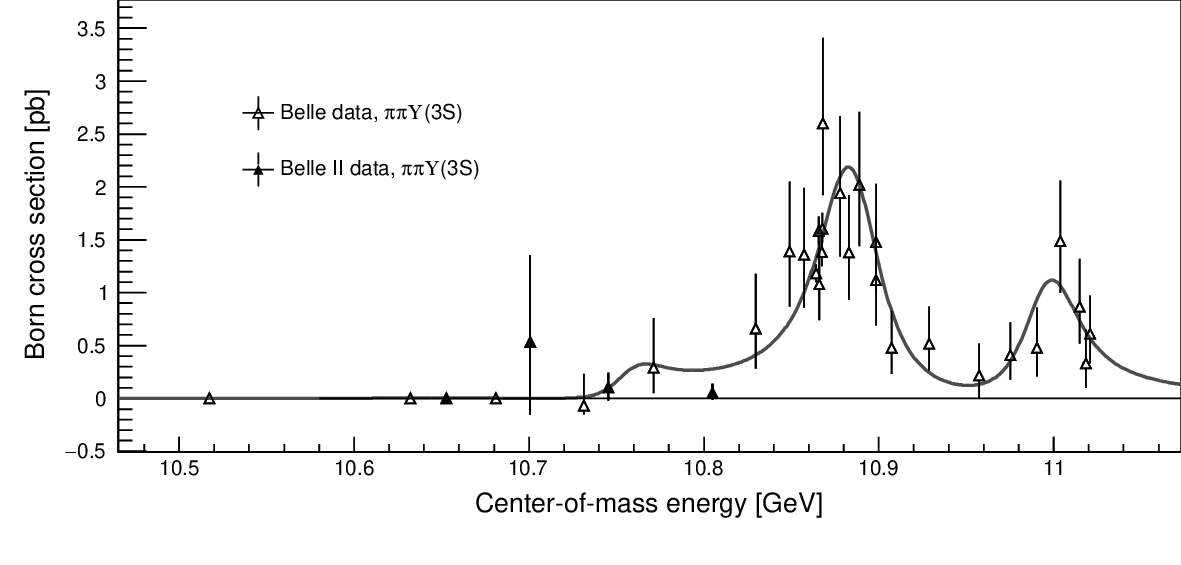}    
  \caption{Born cross sections for $\pi\pi\Upsilon(1S)$ (top), $\pi\pi\Upsilon(2S)$ (middle), and $\pi\pi\Upsilon(3S)$ (bottom), with fit results overlaid. Points with error bars show measured cross sections, solid curves are the results of the simultaneous fit results.}
  \label{fig:xsfit_separate}
\end{figure}

After including the systematic uncertainties that are discussed in detail later, we fit the Born cross section as a function of $\sqrt{s}$ including values from the previous Belle measurement~\cite{exp_belle} with three interfering Breit-Wigner functions representing the $\Upsilon(10753)$, $\Upsilon(5S)$, and $\Upsilon(6S)$ states,
\begin{equation}
\sigma \propto \left|\sum_{i}^{3} \frac{\sqrt{12\pi\Gamma_i\mathcal{B}_i}}{s-M_i+iM_i\Gamma_i} \cdot\sqrt{\frac{f(\sqrt{s})}{f(M_i)}} e^{i\phi_i} \right|^2 \otimes G(0,\delta E),
\end{equation}
where $M_i$, $\Gamma_i$, $\mathcal{B}_i$, and $\phi_i$ are the mass, width, relative branching fraction to $\pipi\Upsilon(nS)$, and relative phase of the $i$th resonance, respectively.
The parameter $f$ is the integral of the three-body phase space at the relevant energy,
and $\otimes G(0,\delta E)$ represents convolution with a Gaussian function used to model the collision energy spread, $\delta E\equiv 5.6~\rm MeV$.
All parameters of the Breit-Wigner functions are free in the fit.

We weight the simulated signal samples according to the fitted cross-section lineshape to make the simulated data better describe real data. 
The weight of the $j{\rm th}$ event generated at the $i{\rm th}$ energy is calculated by the following formula:
\begin{equation}\label{Fun:Wight}
    w_{i,j}= \frac{\sigma^{\rm new}(\sqrt{s}=M^{\rm gen}_{i,j}(\Upsilon(10753)))} {\sigma^{\rm ini}(\sqrt{s}=M^{\rm gen}_{i,j}(\Upsilon(10753)))}, 
\end{equation}
where $\sigma^{\rm new}(\sqrt{s})$ is the Born cross section calculated with the fitted cross-section lineshape, while $\sigma^{\rm ini}(\sqrt{s})$ is the initial cross-section lineshape. 
The value of $M_{i,j}^{\rm gen}(\Upsilon(10753))$ is the generator-level invariant 
mass of the parent particle $\Upsilon(10753)$ for the $j$th event generated at the $i$th energy in 
the initial simulated sample before any selection.

After obtaining the weighted simulation samples, the weighted efficiency $\epsilon$ is calculated by the following formula:
\begin{equation}\label{Fun:efficiency}
    \epsilon^{\rm wtd}_{i}= \frac{\sum^{N_{\rm rec}}_{k=0} w_{ik}} {\sum^{N_{\rm gen}}_{j=0} w_{ij}}, 
\end{equation}
where $N_{\rm rec}$ is the number of events remaining after the event selection at the $i{\rm th}$ energy point, and $N_{\rm gen}$ is the generated number of the simulated signal events.  
The simulated signal lineshapes used in the fit are also updated with the resulting weights.
The Born cross sections are then re-calculated and re-fitted, and this process is repeated. 
Stable results for the correction factors and Born cross sections are obtained within five iterations.
The final results are summarized in Tab.~\ref{tab:born_results}.
We determine the statistical significance of the $\Upsilon(nS)$ signals using their likelihood ratios relative to the background-only hypothesis.
The c.m.\ energies are calibrated using the $\epem\to\B^{(*)}\bar B^{(*)}$
processes~\cite{Belle-II:BBscan}; corresponding values are also shown
in Tab.~\ref{tab:born_results}.

The Born cross sections and the fit to their energy dependence are displayed in Fig.~\ref{fig:xsfit_separate}.
Clear signals of the $\Upsilon(10753)$ state are seen in the $\pi\pi\Upsilon(1S)$ and $\pi\pi\Upsilon(2S)$ channels.
We first fit the Born cross sections from the three channels individually.
The $\Upsilon(10753)$ mass from the individual $\pi\pi\Upsilon(1S)$ and $\pi\pi\Upsilon(2S)$ fits are found to be $(10758.1\pm5.3)~{\rm MeV}/c^2$ and $(10756.3\pm3.6)~{\rm MeV}/c^2$, with widths of $(25\pm20)~\rm MeV$ and $(34\pm15)~\rm MeV$, respectively.
The consistency of the mass and width values suggests that the structure found in $\pi\pi\Upsilon(1S)$ and $\pi\pi\Upsilon(2S)$ is the same.
The significances of the $\Upsilon(10753)$ from the $\pipi\Upsilon(1S)$ and $\pipi\Upsilon(2S)$ channels individually are $4.1\sigma$ and $7.5\sigma$, respectively.
In contrast, no decays of the $\Upsilon(10753)$ to $\pi\pi\Upsilon(3S)$ final states are evident.
We fit the Born cross sections from $\pi\pi\Upsilon(3S)$ data with the $\Upsilon(10753)$ parameters fixed to the expected values, and the significance is only $0.2\sigma$.

We then fit the Born cross sections from $\pi\pi\Upsilon(1S)$, $\pi\pi\Upsilon(2S)$, and $\pi\pi\Upsilon(3S)$ channels simultaneously with common resonance mass and width parameters.
The simultaneous fit is shown as the solid curve in Fig.~\ref{fig:xsfit_separate}.
The goodness of fit is $\chi^2/\rm n.d.f.$ = $89.3/70=1.28$, where n.d.f. is the number of degrees of freedom.
The mass and width of the $\Upsilon(10753)$ state are found to be $(10756.6\pm2.7)$ MeV/$c^{2}$ and $(29.0\pm8.8)$ MeV, respectively.
In addition, the parameters of the $\Upsilon(5S)$ are measured to be $(10884.5\pm1.2)$ MeV/$c^2$, and $(38.5\pm3.6)$ MeV, and $\Upsilon(6S)$ are $(10995.8\pm4.2)$ MeV/$c^2$, and $(33.5\pm8.6)$ MeV.
The parameters of the $\Upsilon(5S)$ and $\Upsilon(6S)$ resonances are consistent with the world average values~\cite{pdg}.

We determine the ratios between $\pipi\Upsilon(1S)$ and $\pipi\Upsilon(2S)$ cross sections, and between $\pipi\Upsilon(3S)$ and $\pipi\Upsilon(2S)$ cross sections at three different energies corresponding to the $\Upsilon(10753)$, $\Upsilon(5S)$, and $\Upsilon(6S)$ resonance peaks based on the simultaneous fit result and the covariance matrix.
The results are given in Tab.~\ref{tab:xs_relative_ratios}.
The ratios between $\pipi\Upsilon(1S)$ and $\pipi\Upsilon(2S)$ channels are consistent among the three resonance peaks, while the ratio of the $\pipi\Upsilon(3S)$ and $\pipi\Upsilon(2S)$ channels from $\Upsilon(10753)$ peak is significantly smaller than other two.

\begin{table*}[htb]
  \begin{center}
    \begin{tabular}{l c  c  c  c   c  c }
      \hline 
       &		$\mathcal{R}^{\Upsilon(10753)}_{\sigma(1S/2S)}$  & $\mathcal{R}^{\Upsilon(10753)}_{\sigma(3S/2S)}$   & $\mathcal{R}^{\Upsilon(5S)}_{\sigma(1S/2S)}$  & $\mathcal{R}^{\Upsilon(5S)}_{\sigma(3S/2S)}$     &  $\mathcal{R}^{\Upsilon(6S)}_{\sigma(1S/2S)}$  & $\mathcal{R}^{\Upsilon(6S)}_{\sigma(3S/2S)}$   \\
      \hline
      Ratio & $0.46^{+0.15}_{-0.12}$ & $0.10^{+0.05}_{-0.04}$ & $0.45^{+0.04}_{-0.04}$ & $0.32^{+0.04}_{-0.03}$ & $0.64^{+0.23}_{-0.13}$ & $0.41^{+0.16}_{-0.12}$ \\
\hline
	\end{tabular}
    \caption{Cross-section ratios at resonance peaks above the $\Upsilon(4S)$. Uncertainty in this table combines statistical and systematic uncertainties.}
    \label{tab:xs_relative_ratios}
  \end{center}
\end{table*}

In addition, we set limits on the Born cross sections for the production of $Z_b(10610)^{\pm}$ and $Z_{b}(10650)^{\pm}$.
Assuming these signals originate from $\Upsilon(10753)$ decay, with the latter state forbidden by phase-space at $\sqrt{s}=10.746$ GeV, signal yields are extracted from an extended maximum likelihood fit to $\Delta M_{\pi}$.
As shown in Fig.~\ref{fig:mpipi}, we fit the $\Delta M_{\pi}$ distributions with $Z_b(10610)^{\pm}$ and $Z_b(10650)^{\pm}$ signal components and a phase-space contribution.
Here we restrict the signal yield to be non-negative.
No significant signal is found either for the $Z_b(10610)^{\pm}$ or $Z_b(10650)^{\pm}$. 
Upper limits are estimated with the following method.
By fixing the signal yields over a range of values and allowing the other parameters to vary in the fit, the likelihood value is extracted as a function of the number of $Z_b(10610)$ or $Z_b(10650)$ signal events.
A Gaussian function is convolved with the profile likelihood distribution to approximate the impact of systematic uncertainties, which are described below.
The upper limit on the number of signal events at the 90\% credibility level is the position where the integral area of the distribution equals 90\% of the entire area which integrated starting from zero.
Upper limits on the Born cross sections of $Z_b(10610)^{\pm}$ and $Z_b(10650)^{\pm}$ are calculated using Eq.~\ref{eq:bxs} and the corresponding radiative and polarization factors.
The results are listed in Tab.~\ref{tab:born_results_zb}.

\begin{table}[htb]
 \begin{center}
    \begin{tabular}{l c c c c c c c c }
\hline
Mode & $N_{Z_{b1}}$   & $N^{\rm UL}_{Z_{b1}}$ &  $\sigma_{Z_{b1}}$ (pb)   & $\sigma^{\rm UL}_{Z_{b1}}$ (pb)   &  $N^{\rm UL}_{Z_{b2}}$ & $N_{Z_{b2}}$ & $\sigma_{Z_{b2}}$ (pb) & $\sigma^{\rm UL}_{Z_{b2}}$ (pb) \\
\hline
\multicolumn{2}{l}{$10.746$ GeV} & & & & \\
\hline
$\pi\Upsilon(1S)$ &  $0.0^{+1.6}_{-0.0}$ & $<4.9$  & $0.00^{+0.04}_{-0.00}$  & $<0.13$ &  $-$   & $-$ & $-$   \\
$\pi\Upsilon(2S)$ &  $5.8^{+5.9}_{-4.6}$ & $<13.8$ & $0.06^{+0.06}_{-0.05}$  & $<0.14$ &  $-$   & $-$ & $-$   \\
\hline \multicolumn{2}{l}{$10.805$ GeV} & & &  &\\
\hline
$\pi\Upsilon(1S)$ &  $2.5^{+2.4}_{-1.6}$&  $<5.2$  &  $0.21^{+0.20}_{-0.13}$   & $<0.43$  &  $0.0^{+0.7}_{-0.0}$ & $<5.8$ & $0.00^{+0.03}_{-0.00}$ & $<0.28$    \\
$\pi\Upsilon(2S)$ &  $5.2^{+3.8}_{-3.0}$&  $<12.3$ &  $0.15^{+0.11}_{-0.09}$   & $<0.35$  &  $0.0^{+0.8}_{-0.0}$ & $<6.0$ & $0.00^{+0.04}_{-0.00}$ & $<0.30$ \\
\hline
    \end{tabular} 
    \caption{Signal yields and upper limits at 90\% credibility for $e^+e^-\to\pi Z_b(10610,10650)$, $~Z_b(10610,10650)\to \pi \Upsilon(1S,2S)$ processes and corresponding Born cross-section measurement limits. Uncertainties for the numbers of signal events are statistical only. Here we use $Z_{b1}$ and $Z_{b2}$ as shorthand for $Z_b(10610)$ and $Z_b(10650)$, respectively.}
    \label{tab:born_results_zb}
   \end{center}
\end{table}

\section{Systematic uncertainties}
Sources of systematic uncertainties on the cross sections include tracking and muon-identification efficiency, integrated luminosity, choice of simulated-event generators, trigger efficiency, $\Upsilon(nS)$ branching fractions, the $\pipi$ amplitude, ISR factor, and the fit procedure.
A momentum-dependent tracking uncertainty is obtained from $\bar B^0 \to D^{*+}(\to D^0 \pi^+) \pi^{-}$ and $e^+e^-\to \tau^+\tau^-$ control samples in data, resulting in uncertainties ranging between 1.5\% to 8.4\% per track with lower momentum.
For the tracking with relatively high momentum, i.e., greater than 200 MeV/c, the efficiency uncertainty substantially improves to as low as 0.3\%.
The muon identification uncertainty, obtained from $J/\psi$ decays, dimuon, and two-photon processes, ranges from 0.5\% to 5.3\%.
The uncertainty in the integrated luminosity is 0.7\% while the uncertainty from the choice of generator is about 1.5\%~\cite{Kuraev:1985hb}.
From studies using dimuon and $e^{+}e^{-}\to \pi^{+}\pi^{-}\pi^{+}\pi^{-}\pi^{0}\pi^{0}$ processes, we assign a systematic uncertainty of $1\%$ due to the trigger modeling.
The uncertainty in the branching fraction of $\Upsilon(nS)\to \mu^+\mu^-$ is taken from Ref.~\cite{pdg}.
We enlarge the $\Delta M$ fit range and change the background parameterization from a linear function to a uniform or a quadratic function, taking the small differences in the signal yields between the nominal fit and the alternative fit as the systematic uncertainty on fit modeling.
To estimate the uncertainty related to the signal probability density function parameterization, we use two alternative fit functions: a Gaussian and a Crystal Ball. The parameters are determined from fits to the simulated signal sample. We find these alternatives lead to negligible change in the signal yield compared to that from the nominal fit.

We consider possible bias introduced in weighting the simulated signal sample with the $M(\pipi)$ distribution from the fit result.
For the $\epem\to\pipi\Upsilon(1S)$ mode, since we cannot separate the uniform $M(\pipi)$ distribution and other hypotheses in the fit, the differences between the efficiencies from the default simulation and weighted efficiencies according to the fit are taken as systematic uncertainties.
For the $\epem\to\pipi\Upsilon(2S)$ mode, we vary the weights by their one standard deviation uncertainties from the fit, and take the largest changes as systematic uncertainties.
To test for any dependence of the iterative process on initial conditions, we also try the dimuon cross-section lineshape as a starting alternative; the final cross sections and numerical results are unchanged.
We vary the $\Upsilon(10753)$ parameters that we are using in the calculation of the ISR factors, the changes of the final Born cross-sections are taken as the systematic uncertainties.
The numerical results are summarized in Table~\ref{tab:born_sys}.

For the measurement of the $\Upsilon(10753)$ mass and width, we consider various systematic effects of the lineshape parameterization.
We adopt a procedure similar to that described in the Belle analysis~\cite{exp_belle}, multiplying the width of $\Upsilon(10860)$ by an energy-dependent factor,
\begin{equation}
1-x-y+ x\left(p_1/p_1^{(0)}\right)^3+y\left(\frac{2}{3}p_2/p_2^{(0)}+\frac{1}{3}p_3/p_3^{(0)}\right),
\end{equation}
where $p_1$, $p_2$, and $p_3$ are momenta of the child particles in the $B_s^{*}\bar B_s^*$, $Z_b(10610)\pi$, and $Z_b(10650)\pi$ systems, respectively, and the superscript (0) denotes a momentum calculated for the nominal $\Upsilon(10860)$ mass.
The factors $\frac{2}{3}$ and $\frac{1}{3}$ roughly correspond to the ratio of the $B\bar B^* \pi $ and $B^*\bar B^* \pi $ cross sections in the molecular interpretation of $Z_b(10610)$ and $Z_b(10650)$.
We experimentally set the weights $x$ and $y$ to the values 0.0, 0.2, 0.4 and 0.6 with the restriction $x+y \leq 0.8$.
The largest changes of $\Upsilon(10753)$ parameters, $\pm0.4$ MeV/$c^2$ and $\pm1.0$ MeV for the mass and width, respectively, are taken as systematic uncertainties.
Since the Dalitz plot of $\Upsilon(10753)$ is different from $\Upsilon(5S)$, we vary the coherent fraction of the $\Upsilon(10860)$ from 100\% to 80\% according to the non-$Z_{b}$ fractions in Ref.~\cite{Belle:2014vzn}, and the largest change of $\Upsilon(10753)$ parameters are $\pm0.5$ MeV/$c^2$ and $\pm0.4$ MeV.
To consider the contribution from the tails of $\Upsilon(2S)$ and $\Upsilon(3S)$, we add a coherent constant amplitude in the fit function, and the change of the $\Upsilon(10753)$ mass and width are $0.5~{\rm MeV}/c^2$ and $0.5~{\rm MeV}$, respectively.
In the mass measurement, we include the correlated systematic uncertainty in c.m. energy of 0.5 MeV.
Adding the systematic contributions in quadrature, we obtain systematic uncertainties of $\pm0.9$ MeV/$c^2$ and $\pm1.2$ MeV for the mass and width, respectively.

\begin{table}[h]
\small
  \begin{center}
    \begin{tabular}{l c c c c c c c c c c }
      \hline Mode & $\mathcal{B}$ & $\mathcal{L}$ & Tracking & $\mu$-ID & Trigger & Generator & $M(\pipi)$ & Fit & ISR & Sum  \\
        \hline \multicolumn{2}{l}{$10.653$ GeV} & & & & \\
        \hline
        $\pi\pi \Upsilon(1S)$ & 1.6 & 0.7  & 1.5  &  0.6 &  1.0 & 1.4 &  1.3 &  29.4 &  6.0 & 30.1  \\
        $\pi\pi \Upsilon(2S)$ & 8.8 & 0.7  & 3.1  &  2.0 &  1.0 & 1.4 &   -  &  40.0 & 2.1 & 41.2 \\
        \hline \multicolumn{2}{l}{$10.701$ GeV} & & &  &\\
        \hline
        $\pi\pi \Upsilon(1S)$ & 1.6 & 0.7  & 1.4 &  0.5 & 1.0 & 1.4 &   6.7  &  180.0 & 1.0 & 180.1 \\
        $\pi\pi \Upsilon(2S)$ & 8.8 & 0.7  & 2.9 &  1.5 & 1.0 & 1.4 &    -   &   30.0 & 1.0 &  31.5 \\
        $\pi\pi \Upsilon(3S)$ & 9.6 & 0.7  & 8.4 &  4.5 & 1.0 & 1.4 &    -   &  100.0 & 2.3 & 100.5\\
        \hline
        \multicolumn{2}{l}{$10.746$ GeV} & & & & \\
        \hline
        $\pi\pi \Upsilon(1S)$ & 1.6 & 0.7  & 1.4 & 0.8 & 1.0 & 1.4 &   1.4 &  0.5  &  4.0 &  5.0 \\
        $\pi\pi \Upsilon(2S)$ & 8.8 & 0.7  & 2.6 & 1.9 & 1.0 & 1.4 &     -  &  6.3  &   2.8 & 11.7\\
        $\pi\pi \Upsilon(3S)$ & 9.6 & 0.7  & 7.7 & 4.9 & 1.0 & 1.4 &     - &  73.0 & 18.7 & 76.5 \\
        \hline
        \multicolumn{2}{l}{$10.805$ GeV} & & &  &\\
        \hline
        $\pi\pi \Upsilon(1S)$ & 1.6 & 0.7  & 1.4 & 0.8 & 1.0 & 1.4 &  3.2 &  8.2 & 1.0 &   9.2  \\
        $\pi\pi \Upsilon(2S)$ & 8.8 & 0.7  & 2.3 & 2.4 & 1.0 & 1.4 &    -   &  0.6 & 4.0 &  10.3 \\
        $\pi\pi \Upsilon(3S)$ & 9.6 & 0.7  & 6.0 & 5.3 & 1.0 & 1.4 &    -   &  7.7 & 1.0 &  14.8 \\
      \hline
\end{tabular}
\end{center}
\caption{Summary of the systematic uncertainties for the cross section measurement of $\epem\to\pipi\Upsilon(nS)$ process. The symbol ``-" denotes the uncertainties which is negligible. The unit is \% in this table.}
\label{tab:born_sys}
\end{table}

\section{Summary}

In conclusion, we report a measurement of the Born cross sections for the $e^+e^-\to\pi^+\pi^-\Upsilon(nS)$ processes using a $19.6~\rm fb^{-1}$ data sample in the energy region near 10753 MeV from Belle II.
Signals for the $\Upsilon(10753)$ are observed in the cross section as a function of energy for the $e^+e^-\to\pi^+\pi^-\Upsilon(1S)$ and $\pi^+\pi^-\Upsilon(2S)$ channels with greater than 8 standard deviation significance, while no evidence is found in $\pi^+\pi^-\Upsilon(3S)$ events.

Combining these results with the Belle measurement~\cite{exp_belle}, the cross-section ratios $\sigma(\pi^+\pi^-\Upsilon(1S,3S))/\sigma(\pi^+\pi^-\Upsilon(2S))$ at the $\Upsilon(10753)$ resonance peak are determined for the first time.
The results are $0.46^{+0.15}_{-0.12}$ and $0.10^{+0.05}_{-0.04}$ for the $\pipi\Upsilon(1S)$ and $\pipi\Upsilon(3S)$ channels, respectively.
The ratio for $\pipi\Upsilon(1S)$ channel is compatible with the ratios at the $\Upsilon(5S)$ and $\Upsilon(6S)$ resonance peaks.
However, the relative ratio of $\pi^+\pi^-\Upsilon(3S)$ channel at the $\Upsilon(10753)$ peak is about three-to-four times smaller than those at the $\Upsilon(5S)$ and $\Upsilon(6S)$ peaks.
Comparing the ratios with the predictions~\cite{Bai:2022cfz}, the observed cross section for $\Upsilon(10753)\to\pipi\Upsilon(3S)$ is lower than expectation.

No evidence is found that these transitions occur via intermediate $Z_b(10610/10650)^{\pm}$ states.
The dipion mass distribution in $\pi^+\pi^-\Upsilon(2S)$ production is similar to that observed in the $\Upsilon(2S)\to\pi^+\pi^-\Upsilon(1S)$ process and can be described accurately by the $\Upsilon(nS)$ transition amplitude.
These distributions can provide an input for theoretical calculations regarding dipion transitions and their relation to the physical nature of the parent $\Upsilon(10753)$ state.

The mass and width of $\Upsilon(10753)$ are measured to be $(10756.6\pm2.7\pm0.9)$ MeV/$c^2$ and $(29.0\pm8.8\pm1.2)$ MeV, respectively, which is consistent with previous measurements~\cite{exp_belle}.
These results supersede the previous Belle result~\cite{exp_belle}.
This improvement in accuracy provides a more precise basis for theoretical calculations related to the $\Upsilon(10753)$ and resonances in the $\epem\to\pi^{+}\pi^{-}\Upsilon(nS)$ process.

\section{Acknowledgements}
This work, based on data collected using the Belle II detector, which was built and commissioned prior to March 2019, was supported by
Higher Education and Science Committee of the Republic of Armenia Grant No. 
23LCG-1C011
Australian Research Council and Research Grants
No.~DP200101792, 
No.~DP210101900, 
No.~DP210102831, 
No.~DE220100462, 
No.~LE210100098, 
and
No.~LE230100085; 
Austrian Federal Ministry of Education, Science and Research,
Austrian Science Fund
No.~P~31361-N36
and
No.~J4625-N,
and
Horizon 2020 ERC Starting Grant No.~947006 ``InterLeptons'';
Natural Sciences and Engineering Research Council of Canada, Compute Canada and CANARIE;
National Key R\&D Program of China under Contract No.~2022YFA1601903,
National Natural Science Foundation of China and Research Grants
No.~11575017,
No.~11761141009,
No.~11705209,
No.~11975076,
No.~12135005,
No.~12150004,
No.~12161141008,
and
No.~12175041,
and Shandong Provincial Natural Science Foundation Project~ZR2022JQ02;
the Czech Science Foundation Grant No.~22-18469S;
European Research Council, Seventh Framework PIEF-GA-2013-622527,
Horizon 2020 ERC-Advanced Grants No.~267104 and No.~884719,
Horizon 2020 ERC-Consolidator Grant No.~819127,
Horizon 2020 Marie Sklodowska-Curie Grant Agreement No.~700525 ``NIOBE''
and
No.~101026516,
and
Horizon 2020 Marie Sklodowska-Curie RISE project JENNIFER2 Grant Agreement No.~822070 (European grants);
L'Institut National de Physique Nucl\'{e}aire et de Physique des Particules (IN2P3) du CNRS
and
L'Agence Nationale de la Recherche (ANR) under grant ANR-21-CE31-0009 (France);
BMBF, DFG, HGF, MPG, and AvH Foundation (Germany);
Department of Atomic Energy under Project Identification No.~RTI 4002,
Department of Science and Technology,
and
UPES SEED funding programs
No.~UPES/R\&D-SEED-INFRA/17052023/01 and
No.~UPES/R\&D-SOE/20062022/06 (India);
Israel Science Foundation Grant No.~2476/17,
U.S.-Israel Binational Science Foundation Grant No.~2016113, and
Israel Ministry of Science Grant No.~3-16543;
Istituto Nazionale di Fisica Nucleare and the Research Grants BELLE2;
Japan Society for the Promotion of Science, Grant-in-Aid for Scientific Research Grants
No.~16H03968,
No.~16H03993,
No.~16H06492,
No.~16K05323,
No.~17H01133,
No.~17H05405,
No.~18K03621,
No.~18H03710,
No.~18H05226,
No.~19H00682, 
No.~22H00144,
No.~22K14056,
No.~22K21347,
No.~23H05433,
No.~26220706,
and
No.~26400255,
the National Institute of Informatics, and Science Information NETwork 5 (SINET5), 
and
the Ministry of Education, Culture, Sports, Science, and Technology (MEXT) of Japan;  
National Research Foundation (NRF) of Korea Grants
No.~2016R1\-D1A1B\-02012900,
No.~2018R1\-A2B\-3003643,
No.~2018R1\-A6A1A\-06024970,
No.~2019R1\-I1A3A\-01058933,
No.~2021R1\-A6A1A\-03043957,
No.~2021R1\-F1A\-1060423,
No.~2021R1\-F1A\-1064008,
No.~2022R1\-A2C\-1003993,
and
No.~RS-2022-00197659,
Radiation Science Research Institute,
Foreign Large-Size Research Facility Application Supporting project,
the Global Science Experimental Data Hub Center of the Korea Institute of Science and Technology Information
and
KREONET/GLORIAD;
Universiti Malaya RU grant, Akademi Sains Malaysia, and Ministry of Education Malaysia;
Frontiers of Science Program Contracts
No.~FOINS-296,
No.~CB-221329,
No.~CB-236394,
No.~CB-254409,
and
No.~CB-180023, and SEP-CINVESTAV Research Grant No.~237 (Mexico);
the Polish Ministry of Science and Higher Education and the National Science Center;
the Ministry of Science and Higher Education of the Russian Federation,
Agreement No.~14.W03.31.0026, and
the HSE University Basic Research Program, Moscow;
University of Tabuk Research Grants
No.~S-0256-1438 and No.~S-0280-1439 (Saudi Arabia);
Slovenian Research Agency and Research Grants
No.~J1-9124
and
No.~P1-0135;
Agencia Estatal de Investigacion, Spain
Grant No.~RYC2020-029875-I
and
Generalitat Valenciana, Spain
Grant No.~CIDEGENT/2018/020;
National Science and Technology Council,
and
Ministry of Education (Taiwan);
Thailand Center of Excellence in Physics;
TUBITAK ULAKBIM (Turkey);
National Research Foundation of Ukraine, Project No.~2020.02/0257,
and
Ministry of Education and Science of Ukraine;
the U.S. National Science Foundation and Research Grants
No.~PHY-1913789 
and
No.~PHY-2111604, 
and the U.S. Department of Energy and Research Awards
No.~DE-AC06-76RLO1830, 
No.~DE-SC0007983, 
No.~DE-SC0009824, 
No.~DE-SC0009973, 
No.~DE-SC0010007, 
No.~DE-SC0010073, 
No.~DE-SC0010118, 
No.~DE-SC0010504, 
No.~DE-SC0011784, 
No.~DE-SC0012704, 
No.~DE-SC0019230, 
No.~DE-SC0021274, 
No.~DE-SC0022350, 
No.~DE-SC0023470; 
and
the Vietnam Academy of Science and Technology (VAST) under Grant No.~DL0000.05/21-23.

These acknowledgements are not to be interpreted as an endorsement of any statement made
by any of our institutes, funding agencies, governments, or their representatives.

We thank the SuperKEKB team for the excellent operation of the accelerator and their special efforts to accomplish the center-of-mass energy scan that made these results possible;
the KEK cryogenics group for the efficient operation of the detector solenoid magnet;
the KEK computer group and the NII for on-site computing support and SINET6 network support;
and the raw-data centers at BNL, DESY, GridKa, IN2P3, INFN, and the University of Victoria for off-site computing support.
E. Won is partially
supported by the NRF grant 2022R1A2B5B02001535 and
J. H. Yin and E. Won are by 2019H1D3A1A01101787.

\bibliographystyle{apsrev4-2}
\bibliography{references}

\clearpage

\end{document}